\documentclass[fleqn,usenatbib]{mnras}

\usepackage[T1]{fontenc}
\usepackage{ae,aecompl}

\usepackage{graphicx}	
\usepackage{amsmath}	
\usepackage{amssymb}	
\usepackage{lmodern}


\newcommand{\es}[1]{{{#1}}}
\newcommand{\ess}[1]{{{#1}}}
\newcommand{\rev}[1]{{{#1}}}

\renewcommand{\vec}[1]{\boldsymbol{#1}}
\newcommand{\av}[1]{\left<#1\right>}
\newcommand{\rom}[1]{{#1}}
\newcommand{\fpath}{.}

\title[3D Resistive Reconnection]{MHD simulations of three-dimensional Resistive Reconnection in a cylindrical plasma column}

\author[E. Striani et al.]{
E. Striani,$^{1}$\thanks{E-mail: edoardo.striani@to.infn.it}
A. Mignone,$^{2}$
B. Vaidya$^{2}$
G. Bodo,$^{3}$
A. Ferrari,$^{1,2}$
\\
$^{1}$INFN Torino, via P. Giuria 1, I-10125, Torino, Italy\\
$^{2}$Dipartimento di Fisica, via P. Giuria 1, I-10125, Torino, Italy\\
$^{3}$Osservatorio Astronomico di Torino
}

\date{Accepted XXX. Received YYY; in original form ZZZ}

\pubyear{2016}

\begin{document}
\label{firstpage}
\pagerange{\pageref{firstpage}--\pageref{lastpage}}
\maketitle

\begin{abstract}
Magnetic reconnection is a {plasma} phenomenon where
a topological rearrangement of magnetic field lines with opposite
polarity results in dissipation of magnetic energy
into heat, kinetic energy and particle acceleration.
Such a phenomenon is  considered as an efficient mechanism for
energy release in laboratory and astrophysical plasmas.
{An important question is how to make the process fast enough to account
for observed explosive energy releases.}
The classical model for steady state magnetic reconnection predicts reconnection times scaling as $S^{1/2}$ (where $S$ is the Lundquist number) and
yields times scales several order of magnitude larger than the observed ones. 
Earlier two-dimensional MHD simulations showed that for large Lundquist number the reconnection time becomes independent of $S$ (``fast reconnection'' regime)
due to the presence of the secondary tearing instability that takes place for $S \gtrsim 1 \times 10^4$.
We report on our 3D MHD simulations of magnetic reconnection
in a magnetically confined cylindrical plasma column under
either a pressure balanced or a force-free equilibrium and compare the results with 2D simulations of a circular current sheet.
We find that the 3D instabilities acting on these configurations result in a fragmentation of the initial current sheet in
small filaments, \rev{leading to enhanced dissipation rate that becomes independent of the Lundquist number already at $S \simeq 1\times 10^3$.}
\end{abstract}




\section{Introduction}
%
%
%
Magnetic reconnection is \ess{a plasma } phenomenon where
a rapid rearrangement of magnetic fields of opposite polarity leads to the dissipation
of the magnetic energy into heat, plasma kinetic energy and particle acceleration.
\ess{In particular}, magnetic reconnection is generally
regarded as a mechanism to account for the fast (i.e. much shorter than the dynamical time-scale)
and intense variability observed in many astrophysical environments,
like active galactic nuclei \citep{Giannios2013} and pulsar wind nebulae \citep{Cerutti2013}.
It is also \ess{likely to occur} in space environments like 
solar flares and coronal mass ejection  \citep{Gordovskyy2010a,Gordovskyy2011,Drake2006}.
A measure of the conversion of magnetic energy into particle acceleration via magnetic reconnection
in Earth's magnetosphere is reported in a recent paper of \cite{Burch2016}.
Finally, magnetic reconnection is responsible for sawtooth crashes that prevent the magnetic
confinement in laboratory fusion experiments, such as tokamaks \citep{Hastie1997}.
The general features of steady state magnetic reconnection are described by the theory of Sweet-Parker \citep{Sweet1958,Parker57},
that \ess{proposed reconnection taking place in current sheets} 
(\rev{localized} regions of very intense currents where non-ideal effects become important) of length $L$ and thickness $\delta$.
In this model the reconnection time scales as
$S^{1/2}$ (where $S = LV_{A}/\eta$, is the Lundquist number, $L$ is the
characteristic length of the field configuration, $V_{A}$ is the
Alfv\'en velocity and $\eta$ is the resistivity).
However, considering that the Lundquist number is very large in space, astrophysical and laboratory plasmas
\citep[e.g. $S \sim 10^{12}\-- 10^{14}$ in the solar corona and
$S \sim 10^6 \-- 10^8$ in tokamaks, see ][]{Loureiro2016}, the above mentioned
scaling yields reconnection time-scales that are several order
of magnitudes longer than observed.
An attempt to solve this problem was suggested by Petschek \citep{Petschek1964},
whose model yields a logarithmic dependence of the reconnection rate on $S$.
\ess{Petschek-like configuration and scaling are found in a recent relativistic resistive 
magnetohydrodynamic (MHD) simulation of \cite{DelZanna2016}. However, this regime was never observed in laboratory experiments. 

The understanding of this time-scale problem was significantly improved using resistive MHD numerical simulations with large Lundquist number.
Two-dimensional simulations \citep[see, e.g., ][]{ Samtaney2009,Loureiro2012,Huang2010,Huang2013}
have shown that when  $S > S_c\simeq 1 \times 10^4$,}
the current sheet is subject to secondary tearing instability \citep{Biskamp1986}, \ess{resulting in the} fragmentation of the current sheet \ess{and formation of } a large number of plasmoids.
This leads to \ess{the so called} ``fast reconnection'' regime, where the reconnection rate becomes independent of the resistivity.
Recent three-dimensional MHD simulations \citep{Oishi2015} have further shown that 3D instabilities can trigger a ``fast reconnection'' regime even for $S < S_c$.

\begin{figure}
  \centering
 \includegraphics[width=8.1cm]{\fpath/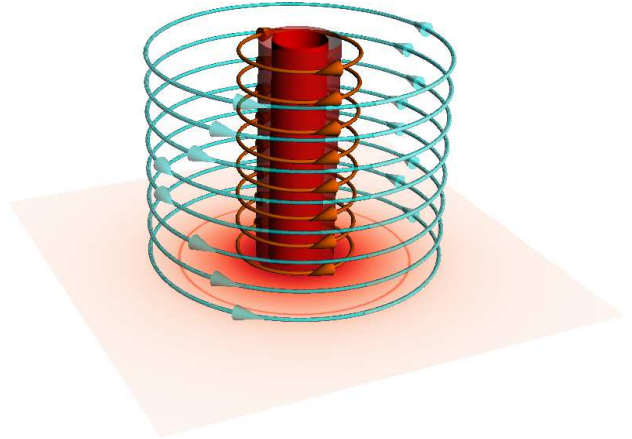}
 \caption{Contour rendering of the density showing the plasma column at $t = 0$ for the pressure balanced equilibrium.
  A slice of the density (pseudocolor rendering ) on the $xy$ plane ($z = -1$) is superimposed,
  showing the maximum of $\rho$ at $r_1$ (see text). The red and blue
 circles show magnetic field lines of opposite polarity for $B_{\phi}$.}\label{Plasma_Column}
\end{figure}

In the present work, we consider both 2D and 3D resistive MHD simulations of magnetically confined cylindrical plasma columns (\es{see Fig. 
\ref{Plasma_Column}}) featuring a current ring where the azimuthal component of magnetic field changes polarity. \rev{This field configuration
was considered in \cite{Romanova1992} and more recently in \cite{McKinney2012}.}
We consider two initial equilibria: one in which radial force balance is established by a thermal pressure gradient and one in which the field is 
force-free. 
The former can become unstable to pressure-driven instabilities while the latter is prone to the onset of current-driven modes.
We then show that the presence of 3D plasma column instabilities results in a fragmentation of the initial current sheet
and leads to a ``fast reconnection'' regime also for $S \simeq 1\times 10^3$.

This paper is organized as follows. In section \ref{Equations} we summarize the equations of resistive MHD
used in the simulations and we present our model setup and initial conditions.
In section \ref{Results} we illustrate the results of our simulations. Finally in section \ref{Discussion}
we summarize and discuss our findings.

\section{Equations and Model Setup}\label{Equations}
%
%
%
We solve the equations of  {resistive} MHD listed below using the
PLUTO code for astrophysical gasdynamics \citep{Mignone2007,Mignone2012}:
\begin{equation}
  \begin{array}{lcl}
    \displaystyle \frac{\partial \rho}{\partial t} + \nabla \cdot (\rho \vec{v}) &=& 0 \\ \noalign{\medskip}
    \displaystyle \rho \left[\frac{\partial \vec{v}}{\partial t} + (\vec{v} \cdot \nabla )\vec{v}\right] + \nabla p  - (\nabla \times \vec{B}) \times \vec{B} &=& 0 \\ \noalign{\medskip}
    \displaystyle \frac{\partial \vec{B}}{\partial t} - \nabla \times (\vec{v} \times \vec{B} - \eta \nabla \times \vec{B}) &=& 0
  \end{array}
\end{equation}
Here $\rho$, $\vec{v}$, $\vec{B}$ and $p$ are, respectively, the fluid mass density, velocity, magnetic field and gas pressure.
Proper closure is given by an
isothermal equation of state, $p = c_s^2\rho$ (where $c_s$ is the isothermal speed of sound).
The equations are solved in conservative form using a second-order Runge-Kutta time stepping with
linear reconstruction and the Riemann solver of Roe \citep{Roe81}.

We consider a magnetized plasma column in which the azimuthal component takes the form
\begin{equation}\label{Eq:Bphi}
B_{\phi}(r) = B_0 \frac{r/a}{1+(r/a)^2}\tanh\left(\frac{r-r_1}{w}\right),
\end{equation}
where $r_1$ is the radius of field inversion and
$w$ is the width of the current sheet.
Equation (\ref{Eq:Bphi}) has a maximum at  $r= a$ where $B_{\phi}^{max} = B_0/2$.

\rev{Our aim is to investigate the evolution of a cylindrical plasma column featuring a field in the form of Eq.(\ref{Eq:Bphi}).
For simplicity, we will start by assuming an initial equilibrium configuration based on radial force balance.
One should be aware, however, that such an equilibrium could not be realized as it is potentially prone to 
many types of instabilities, as we shall see. 
The issue of marginal equilibrium was recently addressed by \cite{Uzdensky2016}.}
Radial force balance is achieved by solving the radial component of the momentum equation, which reads
\begin{equation}\label{Eq:equilibrium}
 \frac{d\Pi}{dr} = - \frac{1}{2r^2}\frac{d}{dr}\left(r^2 B^2_{\phi}\right)
\end{equation}
where $\Pi = p + B_z^2/2$.
Eq. (\ref{Eq:equilibrium}) has solution
\begin{equation}\label{Eq:solution}
 \Pi = \Pi_0 - \left.\frac{B^2_{\phi}}{2}\right|^r_{0} - \int_{0}^r \frac{B^2_{\phi}}{r}\,dr
  \,,
\end{equation}
where the integration constant,
\begin{equation}
  \Pi_0 = p_0 + \left.\frac{B_z^2}{2}\right|_{r=0} \,.
\end{equation}
%
The input parameters are the pitch $P$ and the plasma $\beta$
(a factor $1/\sqrt{4\pi}$ is absorbed in the definition of $\vec{B}$):
\begin{equation}
 P = \left.\frac{r B_z}{B_{\phi}}\right\rvert_{r=0} \,\,\,\,\,\,\,\,,
\,\,\,\,\,\,\,\, \beta = \frac{2p}{|\vec{B}|^2}.
\end{equation}
The plasma $\beta$ is computed as the ratio of the on-axis gas pressure $p_0$ to the maximum $B_{\phi}$ value:
\begin{equation}
\label{eq:beta}
 \beta = \frac{2p_0}{(B_{\phi}^{max})^2} = \frac{8p_0}{B_0^2} \,\,\,\,\,\, \Rightarrow \,\,\,\,\,\, B_0 = \sqrt{\frac{8p_0}{\beta}}
\end{equation}
We employ an isothermal equation of state ($p = \rho c_s^2$) \es{and adopt periodic boundary conditions in the vertical ($z$) direction while 
equilibrium values are prescribed on the remaining sides.}
Lengths are measured in units of $r_1$, velocities in
units of the isothermal sound speed ($c_s$), and $\rho$ in units of the density at the axis.
The computational domain is the Cartesian box with \rev{$x,y\in[-l/2,l/2]$ and $z\in[-l_z/2, l_z/2]$ where \es{$l = 4r_1$}.}
A random perturbation in $v_x$, $v_y$, $v_z$ of amplitude $1\%$ of the sound speed is added.
An additional perturbation comes from the $m = 4$ noise due to the spatial discretization of the cylindrical plasma column on a cartesian grid.

\subsection{\rev{Equilibrium} Balance}
%
%
%

\begin{figure}
  \centering
 \includegraphics[width=8.5cm]{\fpath/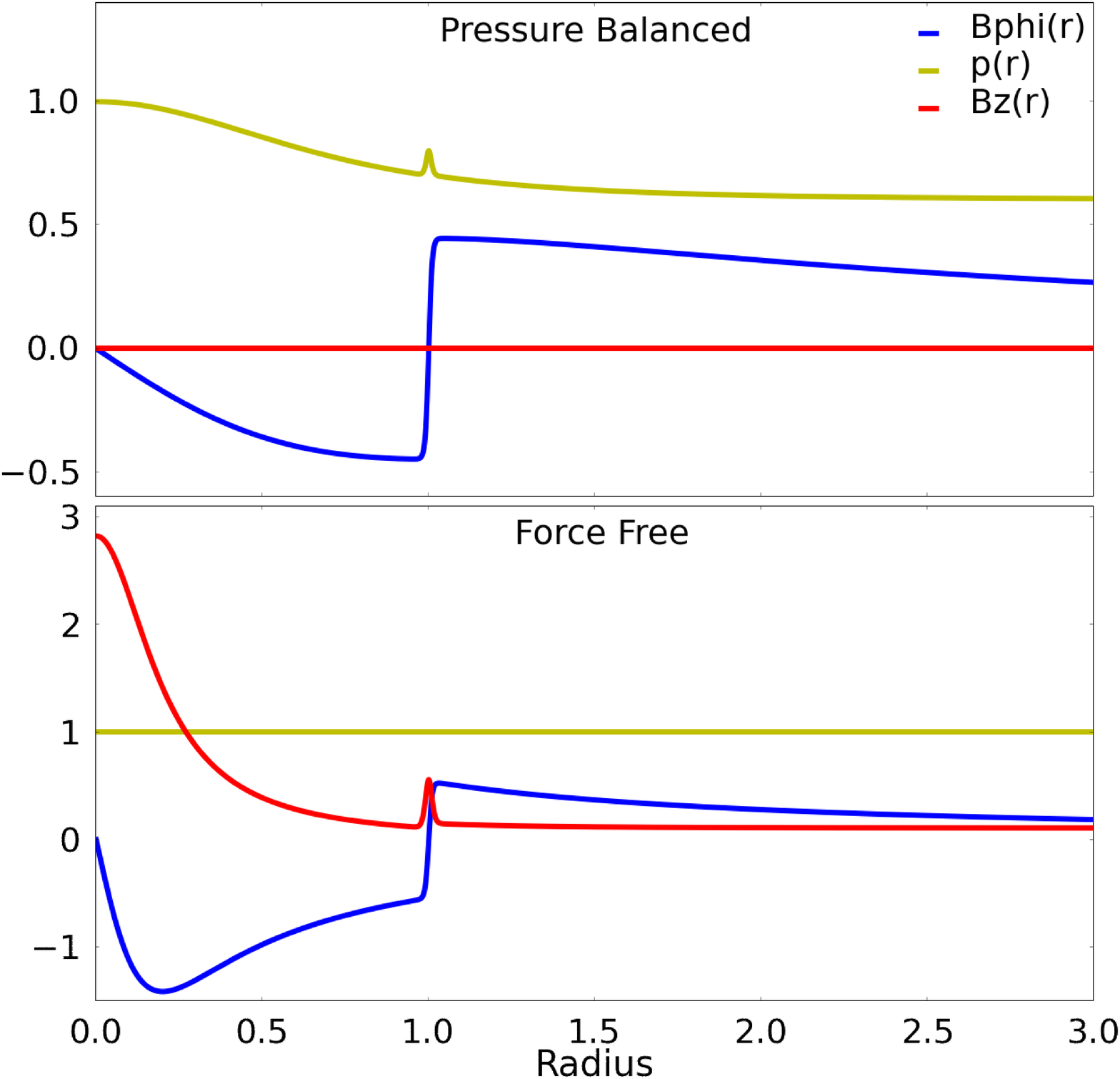}
 \caption{Radial profiles of $B_{\phi}$, $B_z$ and $p$ for the pressure balanced (case PB-0)
 and force-free (case FF-0.2) equilibrium configurations. In both cases the Lundquist number is $S = 2.4 \times 10^4$.}\label{Bphi}
\end{figure}

Two possible equilibrium configurations will be investigated: one in which the Lorentz force
is balanced by a pressure gradient and the other in which the Lorentz force vanishes.
We will refer to the first one as the pressure-balanced (PB) while to the other as force-free (FF).
We set \rev{$l_z=2r_1$} for the PB case while we choose \rev{$l_z = 4r_1$} for the FF case in order to accommodate the kink mode that has the maximum growth at long wavelength.
\es{The radial profiles of $B_{\phi}$, $B_z$ and $p$ for both configurations are shown in Fig. \ref{Bphi}.}
In each of the above configurations we also simulate runs without inversion in toroidal magnetic field, i.e., by setting
the hyperbolic tangent term in eq. \ref{Eq:Bphi} to unity.  These runs without magnetic shear are denoted by the suffix -NS.

\subsubsection{Pressure-Balance (PB) Equilibrium.}
%
%
%
We assume a constant vertical field
\begin{equation}
 B_z(r) = B_0 \frac{P}{a}
\end{equation}
and solve Eq. \ref{Eq:solution} for the pressure $p$:
\begin{equation}\label{eq:pressure}
 p(r) = p_0 - \left.\frac{B^2_{\phi}}{2}\right|^r_{0} - \int_{0}^r \frac{B^2_{\phi}}{r}\,dr
  \,,
\end{equation}
where $p_0 = \rho c_s^2 = 1$ and the integral is solved numerically. This is a variable pitch profile.
\es{Without the magnetic shear the solution of Eq. \ref{Eq:solution} is
\begin{equation}
p(r) = p_0 - \frac{B_0^2}{2}\frac{r^2(2a^2+r^2)}{(a^2+r^2)^2}  \,,
\end{equation}
which has the constraint $p_0/B_0^2 > 1$. From Eq. \ref{eq:beta} this implies $\beta > 4$.}

The growth rates of unstable modes are typically of the order of $c_s /R$,
where $c_s$ is the sound speed and $R$ is the jet radius \citep{Longaretti2008}.

PB equilibria may be subject to pressure-driven instabilities (PDI) driven by perpendicular currents.
They occur in plasma columns when the pressure force pushes the plasma out from the inside of the magnetic field lines curvature and their destabilizing term is proportional to the pressure gradient.
This instability has a very short wavelength perpendicular to the magnetic field and long wavelengths parallel to the field \citep{Freidberg2014}.

\subsubsection{Force-Free (FF) Equilibrium.}
%
%
%
We assume constant pressure $p(r) = p_0$ and solve Eq. \ref{Eq:solution} for the vertical field:
\begin{equation}\label{eq:FF}
  \frac{B_z^2(r)}{2} = \frac{B_{z0}}{2} - \left.\frac{B_{\phi}^2}{2}\right|_0^r - \int_{0}^{r}\frac{B_{\phi}^2}{r}dr.
\end{equation}
Without the magnetic shear, the vertical field has the solution
\begin{equation}\label{eq:Bz}
  B_z(r) = B_0 \sqrt{\frac{P^2(a^2+r^2)^2 - r^2a^2(2a^2+r^2)}{a^2(a^2+r^2)^2}}
\end{equation}
A necessary condition for the \ess{square root } to be positive for $r \to \infty$ is therefore that $P \geq a$.
In our simulations we choose $P = a$.

FF configurations \ess{may be} prone to current-driven instabilities (CDI) driven by parallel currents.
The $ m = 1$ (where $m$ is the azimuthal wavenumber) ``kink'' mode is the most violent among CDI \citep{Begelman1998}.
In this context, three dimensional MHD simulations of relativistic
jets possessing an axial current have shown a prominent jet wiggling due to the growth of non-axial symmetric perturbations \citep[see, e.g.,][]{Mignone2010, Mignone2013}.

\section{Results}\label{Results}
%
%
%
\begin{table*}
  \begin{center}
  \caption{Different 2D and 3D cases with pressure balance (PB) and
  force free (FF) initial conditions along with their setup parameter values.}
  \label{Tab:Cases}
  \begin{tabular}{ l | c | c | c |c | c}
    \hline
    Case 		& Eq  	& Lundquist $S$ ($\times 10^4$) & $P$ 	& $\beta$	& Resolution \\ \hline\hline
    PB-0-2D  		& PB 	& 0.3			   	& 0.0 	& 10	& $512 \times 512$\\
    PB-0-2D  		& PB 	& 1.1   			& 0.0 	& 10	& $1024 \times 1024$\\
    PB-0-2D  		& PB 	& 2.4   			& 0.0 	& 10	& $2048 \times 2048$\\
    PB-0-2D  		& PB 	& 3.4, 5.0, 6.6, 10	  	& 0.0 	& 10	& $4096 \times 4096$\\
    PB-0   		& PB 	& 0.3, 0.7, 1.1		 	& 0.0 	& 10 	& $512 \times 512 \times 256$\\
    PB-0   		& PB 	& 2.4			 	& 0.0 	& 10 	& $1024 \times 1024 \times 512$\\
    PB-0.5		& PB	& 2.4				& 0.5	& 10 	& $1024 \times 1024 \times 512$\\
    PB-0-NS		& PB	& 2.4				& 0	& 10	 &	$1024 \times 1024 \times 512$\\
    FF-0.2		& FF	& 0.3, 0.7, 1.1			& 0.2	& 1 	& $512 \times 512 \times 512$\\
    FF-0.2		& FF	& 2.4				& 0.2	& 1 	& $1024 \times 1024 \times 1024$\\
    FF-10		& FF	& 2.4				& 10	& 1 	& $1024 \times 1024 \times 1024$\\
    FF-0.2-NS		& FF	& 2.4				& 0.2	& 1 	& $1024 \times 1024 \times 1024$\\\hline
  \end{tabular}
  \end{center}
\end{table*}

We consider several simulations characterized by different choices of the Lundquist number $S = L V_A/\eta$ (where $\eta$ is the resistivity, 
\rev{$L$ is the characteristic length of the current sheet and $V_A = B/\sqrt{\rho}$ is the Alfv\'en velocity}), plasma $\beta$, magnetic pitch and equilibrium configurations.
The simulation cases along with their parameters are listed in Table \ref{Tab:Cases} and are: case PB-0-2D (2D circular current sheet with pitch $P = 0$), case PB-0 (3D PB with $P = 0$), case PB-0-NS (same as the previous case, but without shear in magnetic field) case PB-0.5 (3D PB with $P = 0.5$), case FF-0.2 (3D FF with $ P = 0.2$), case FF-0.2-NS (same as the previous case, but without shear in magnetic field) and case FF-10 (3D FF with $ P = 10$).
\ess{The simulation cases without inversion are not expected to dissipate magnetic energy via magnetic reconnection.}
They therefore serve as test cases to ensure that the observed dissipation of magnetic energy in runs with inversion arises from
magnetic reconnection.
\rev{For all of our simulations we fix the width of the current sheet to be $w = 0.01$.}
We set the plasma beta to be $\beta = 10$ for the PB cases and $\beta = 1$ for the FF.
\ess{For each case, we express time in units of the \rev{Alfv\'en} time scale, defined as $t_A = 2\pi r_1/V_A$ where $r_1$ is our unit length and $V_A = \max (|\vec{B}|/\sqrt{\rho}|)$ over the entire computational domain at $t=0$.}

Our simulations for the PB and FF configurations stop at $t = 5.5 t_A$ and $t = 10t_A$ respectively, when the evolution of magnetic reconnection and the instabilities
allowed in each configuration is such that a prominent magnetic dissipation, up to $80 \%$ of the initial magnetic energy, is reached.
The resolutions are chosen so as to ensure that the
numerical resistivity is significantly smaller than the physical resistivity.
\es{A preliminary 2D study was performed in order to find the optimal resolutions.
We plotted the magnetic energy $E_m$ vs $t/t_{A}$
for a given Lundquist number and with different resolutions
($512^2$, $1024^2$, $2048^2$, $4096^2$ and $8192^2$),
and we computed the time at which $E_m$ reaches
$80\%$ of the initial magnetic energy, $t_{80}$. We then choose the lowest resolution among those that
yield the same value (within $10\%$) of $t_{80}$.}

\begin{figure}
  \centering
 \includegraphics[width=8.cm]{\fpath/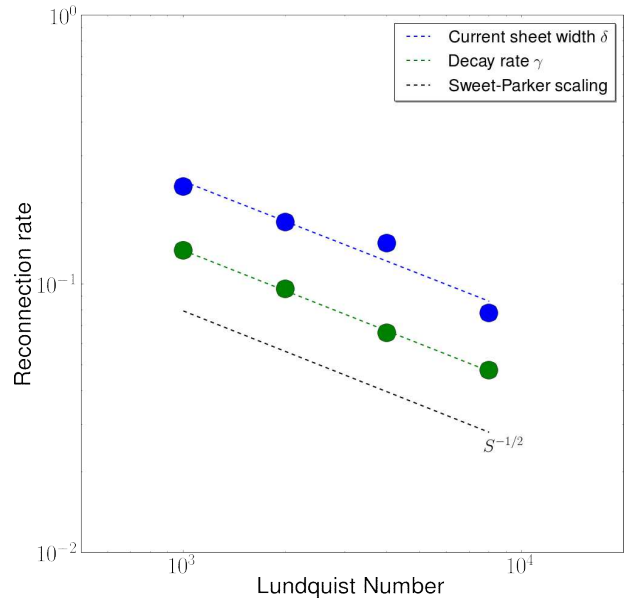}
 \caption{Reconnection rate $\eta$ computed with two different methods:
 \textit{Blue}: time average of $\delta$ as a function of $S$, along with
 a best fit (dashed line).
 \textit{Green}: Magnetic energy decay $\gamma$ as a function of $S$, along with
 a best fit (dashed line).
 The Sweet-Parker scaling $ \sim S^{-1/2}$ is plotted in black.}\label{Fig:Cartesian}
\end{figure}
\begin{figure*}
 \centering
 \includegraphics[height=7.5cm]{\fpath/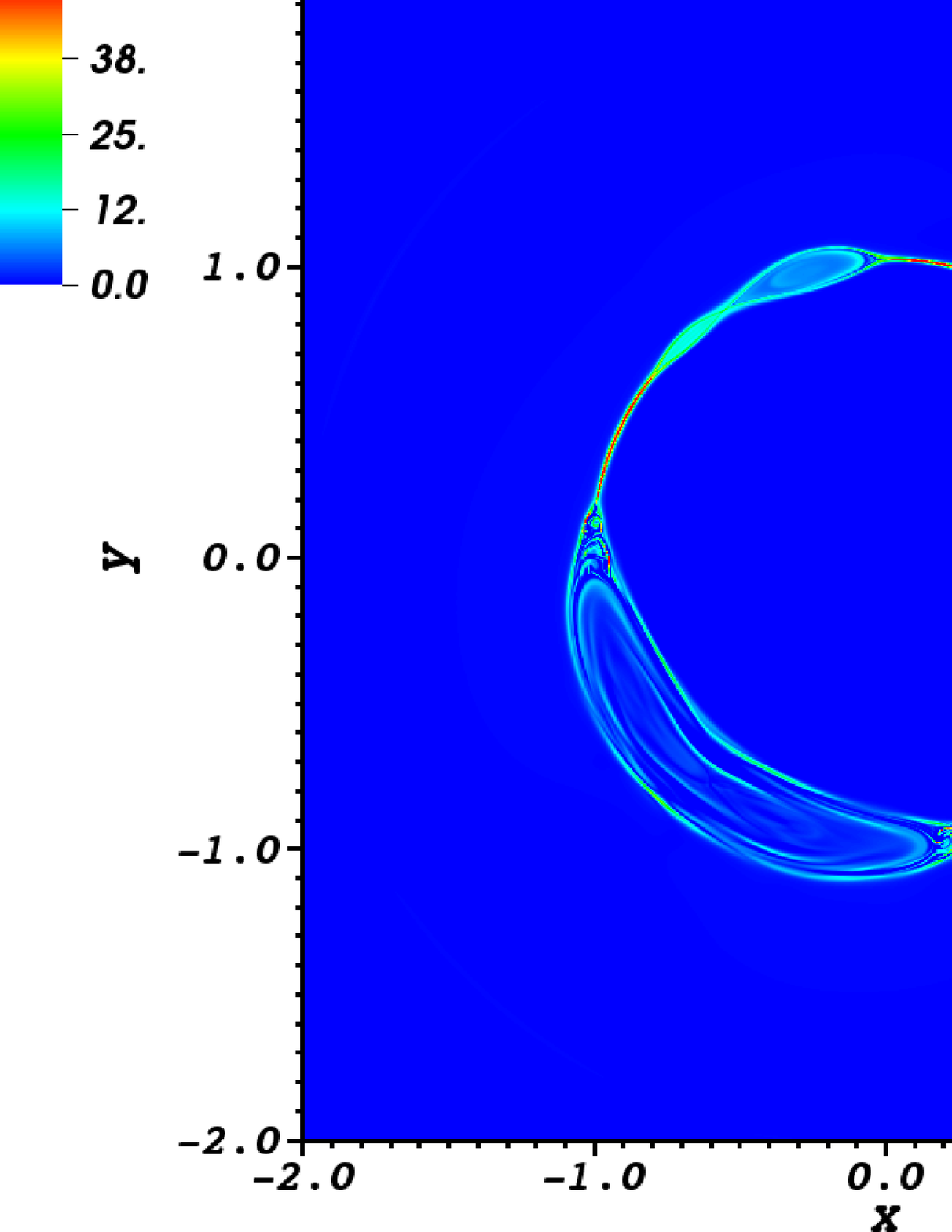}
 \quad
 \includegraphics[height=7.5cm]{\fpath/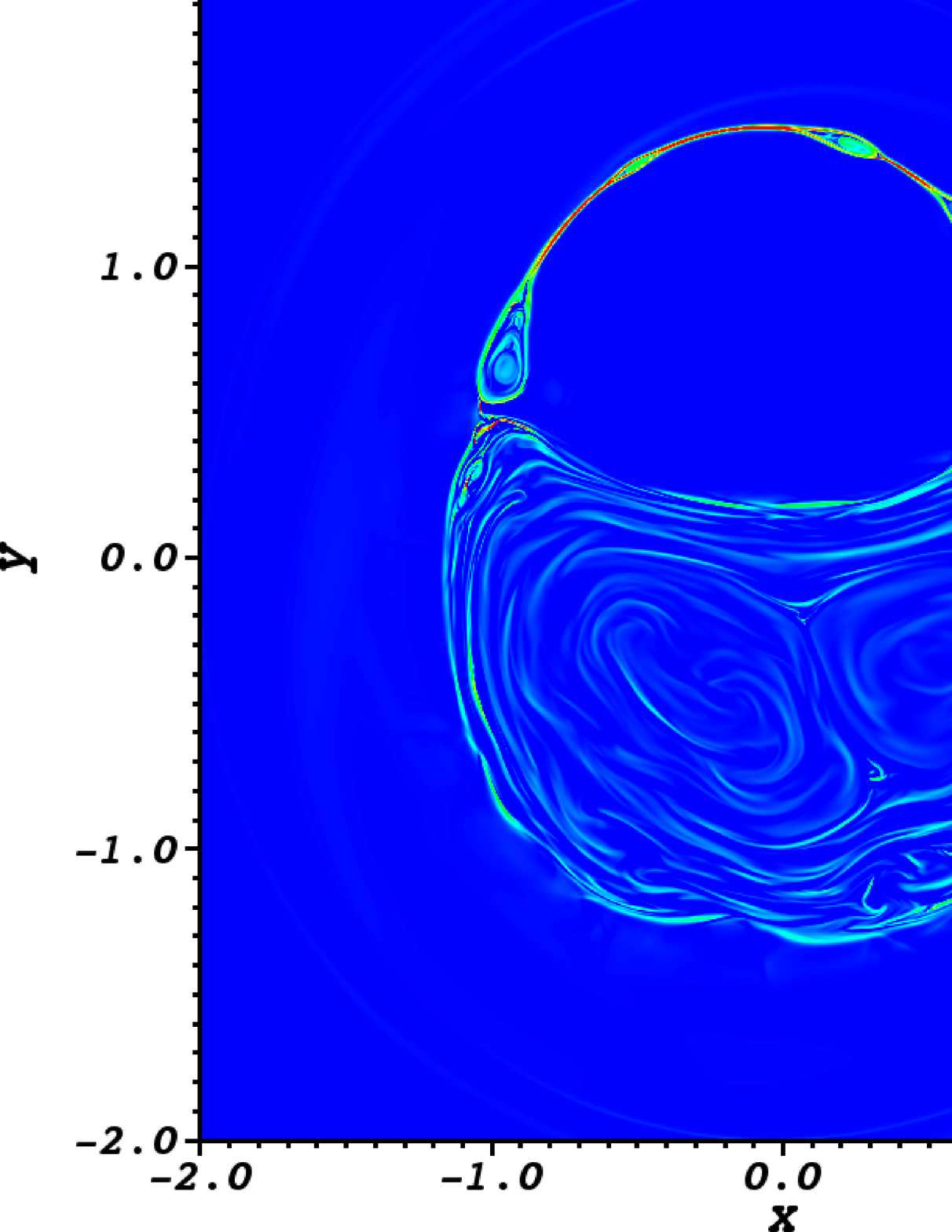}
 \caption{Pseudocolor rendering of the current density $J_{z}$
 for case 2D with $S = 1.0 \times 10^5$ at $t \simeq 2.5 t_{A}$ (left) and
 $t \simeq 4.8 t_{A}$ (right). We note the formation of multiple plasmoids and small-sized current sheets due to secondary tearing instability, 
 in a manner similar to the \rev{``plasmoid turbulence'' described in \protect\cite{Loureiro2012}.}}\label{Fig:2D}
\end{figure*}
\begin{figure}
  \centering
 \includegraphics[width=7.9cm]{\fpath/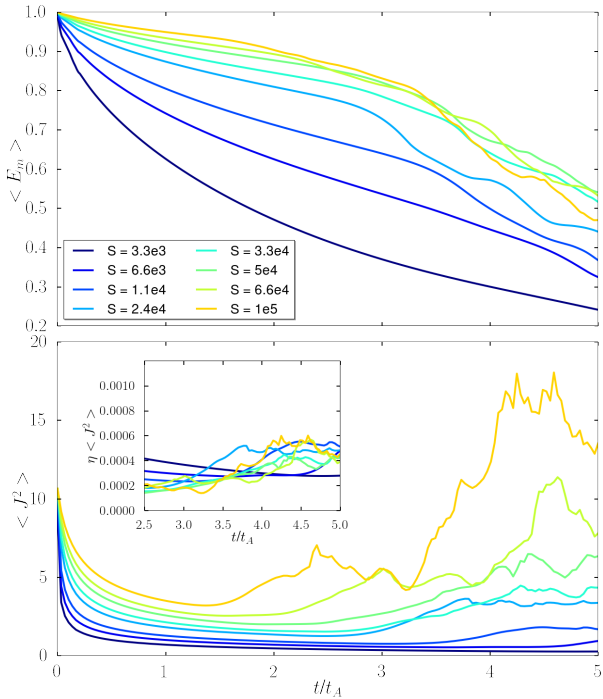}
 \caption{Two-dimensional circular current sheet. \textit{(Top panel)} Temporal evolution of average magnetic energy $E_m$ normalized
 to the initial magnetic energy for different values of the Lundquist number.
 \textit{(Bottom panel)} Temporal evolution of $\av{J^2}$
 for different values of the Lundquist number.
 \rev{\textit{(Inset)} Temporal evolution of $\eta \av{J^2}$ for different Lundquist numbers.}}\label{Fig:2D_Em_J2}
 \end{figure}
\begin{figure}
  \centering
\includegraphics[width=8.cm]{\fpath/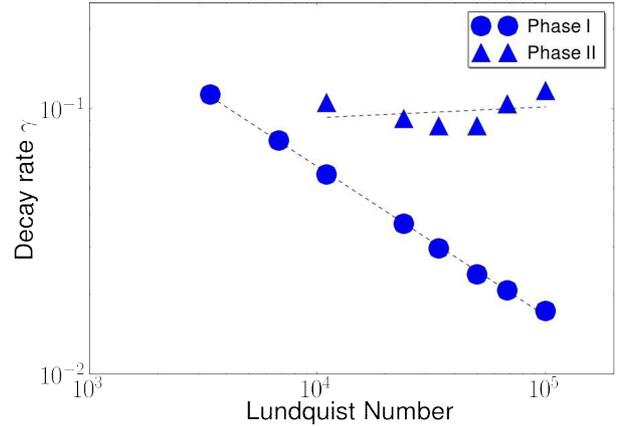}%
 \caption{Decay rate $\gamma$ of the magnetic energy as a function of $S$
 for case PB-0-2D. For each $S$ values
 $\gamma$ is computed in \es{phase I} (circles) and \es{phase \rom 2} (triangles).
 The decay rate scales as $S^{-1/2 }$ during phase I, and it is nearly independent on $S$ in phase II.}\label{Fig:Reconnection_Rate_Polar}
 \end{figure}

\subsection{Two-Dimensional Results}
%
%
%

We begin our discussion by analyzing the two-dimensional case.
For the sake of reference, in \S\ref{sec:harris_sheet} we estimate the reconnection rate in a simple \rev{Harris sheet configuration \citep{Harris1962}}
for later comparison with the actual 2D circular current sheet (\S\ref{Par:Polar2D}).
The 2D circular current sheet is alike case 3D PB-0 case but does not include the $z$ direction.

\begin{figure*}
 \centering
 \includegraphics[width=5.3cm, height = 4.5cm]{\fpath/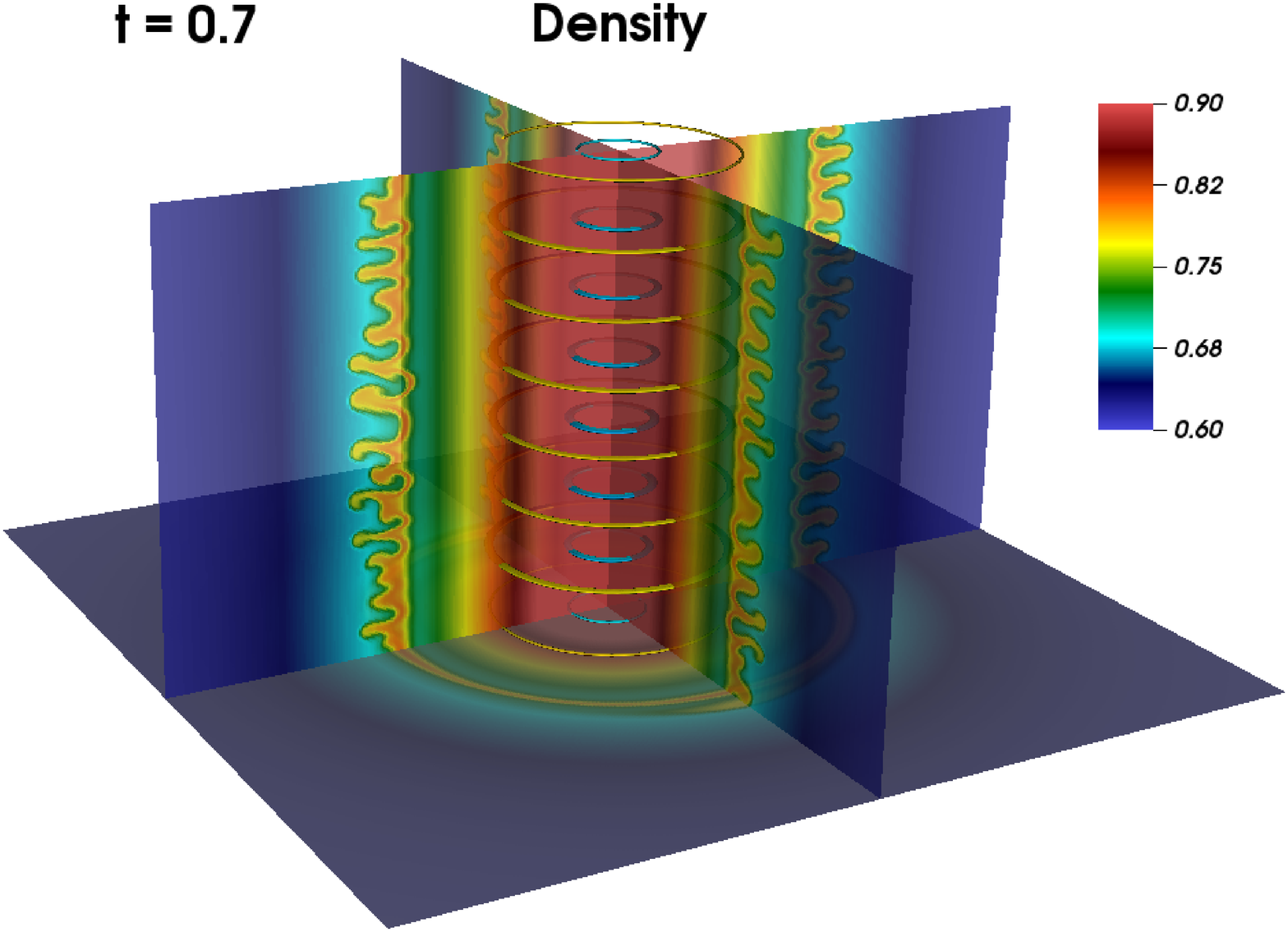}%
 \quad\quad
 \includegraphics[width=5.3cm, height = 4.5cm]{\fpath/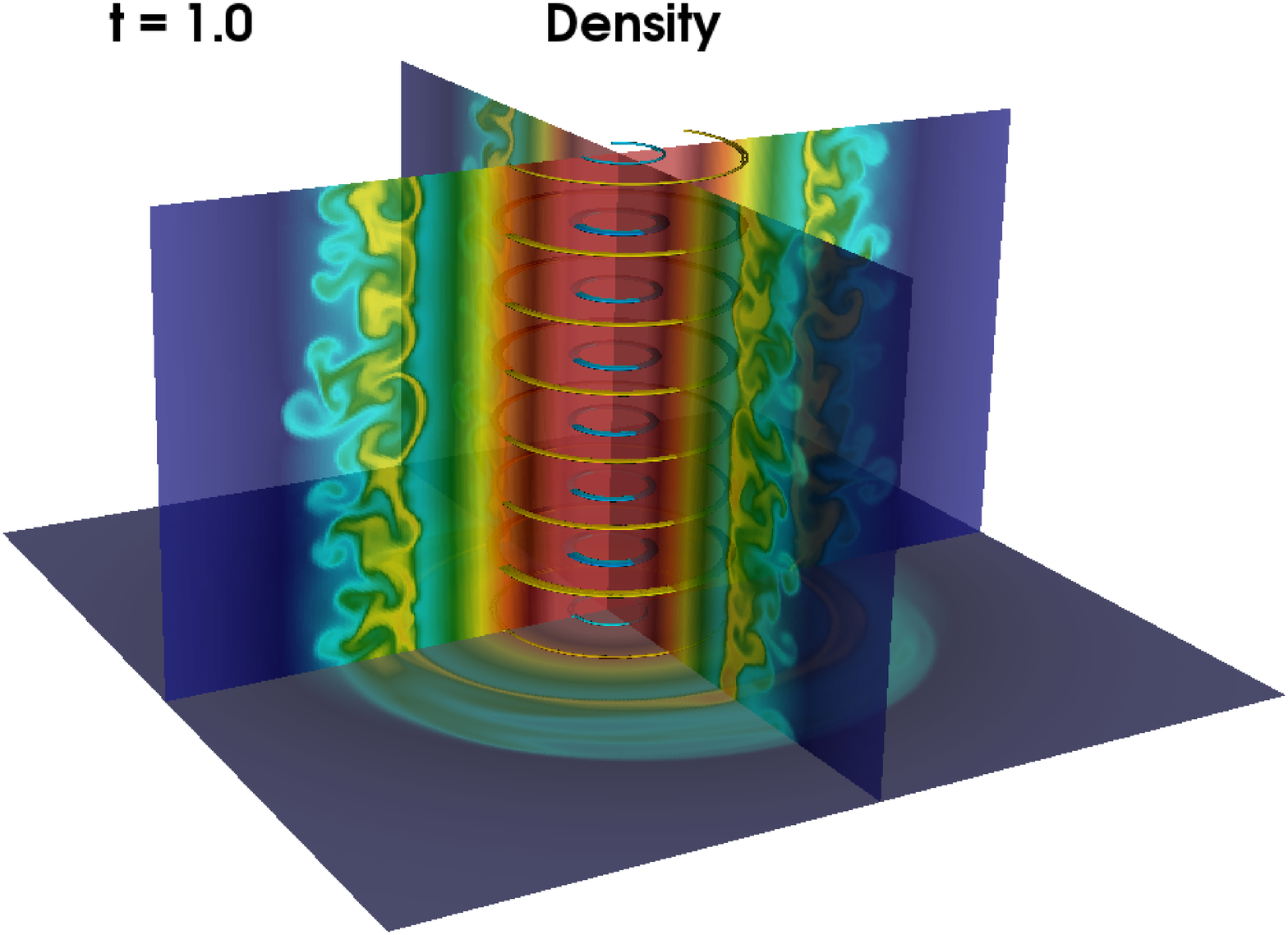}
 \quad\quad
 \includegraphics[width=5.3cm, height = 4.5cm]{\fpath/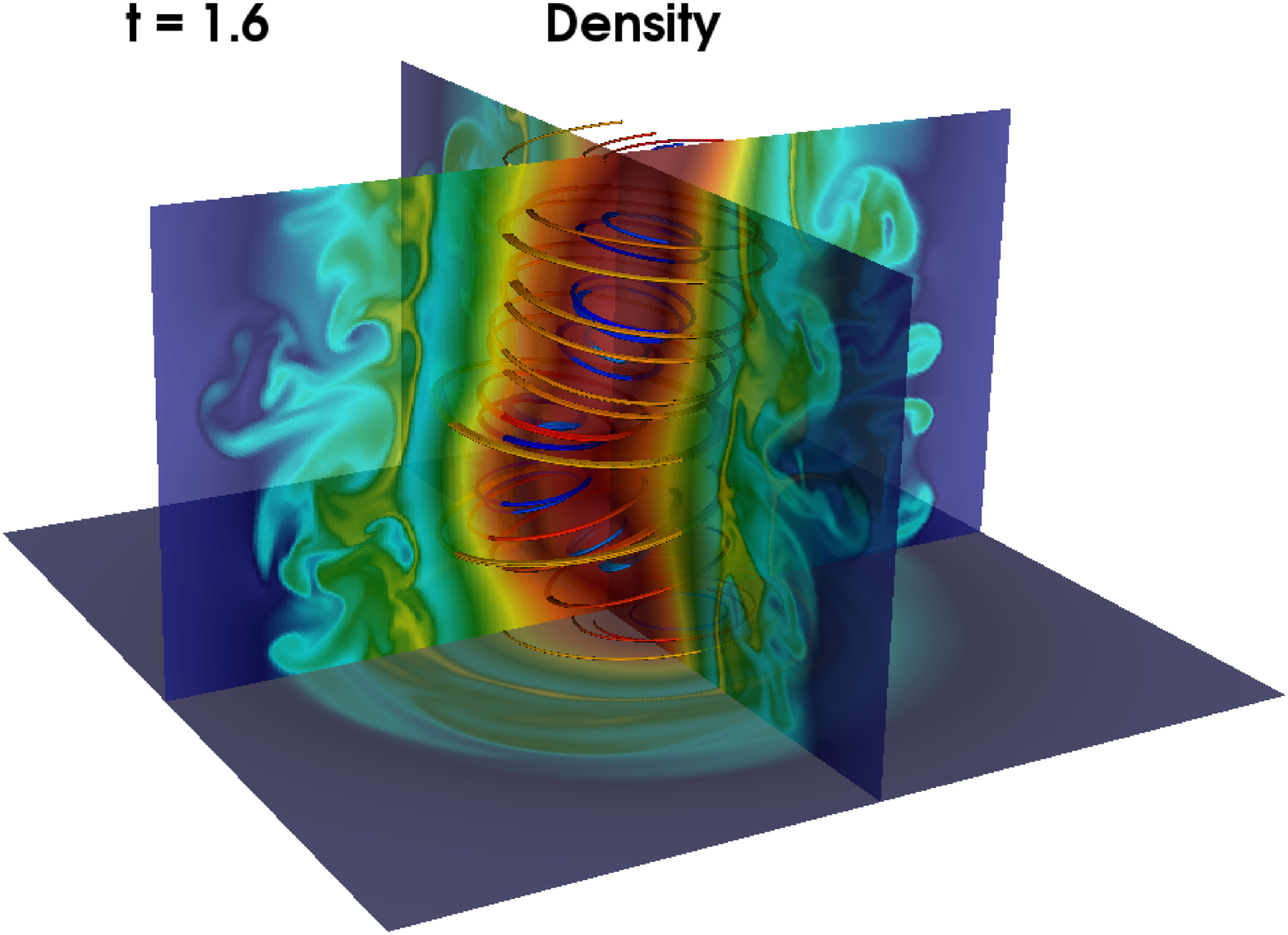}\\
 \vspace{0.5cm}
 \includegraphics[width=5.5cm, height = 3.5cm]{\fpath/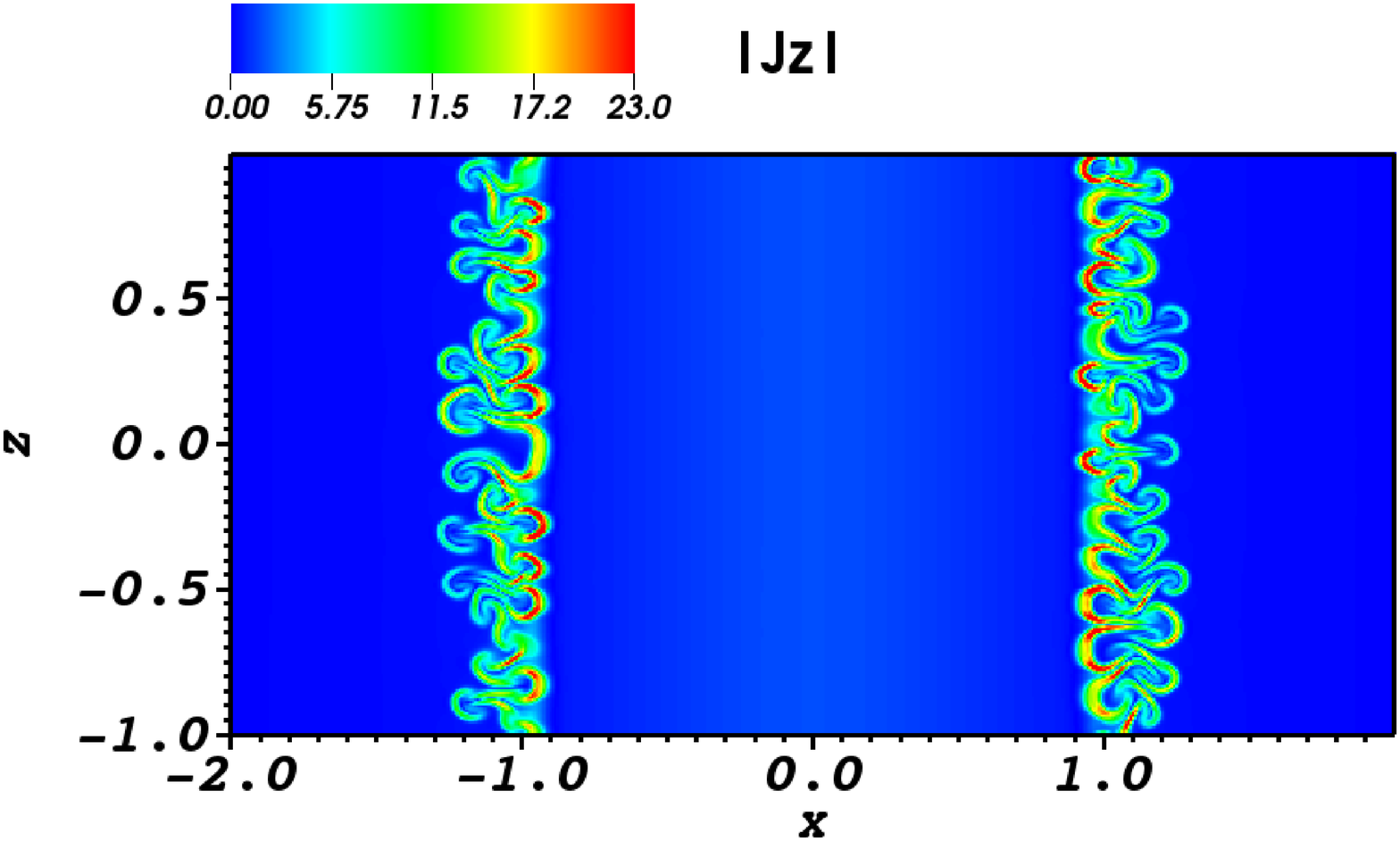}
 \quad
 \includegraphics[width=5.5cm, height = 3.45cm]{\fpath/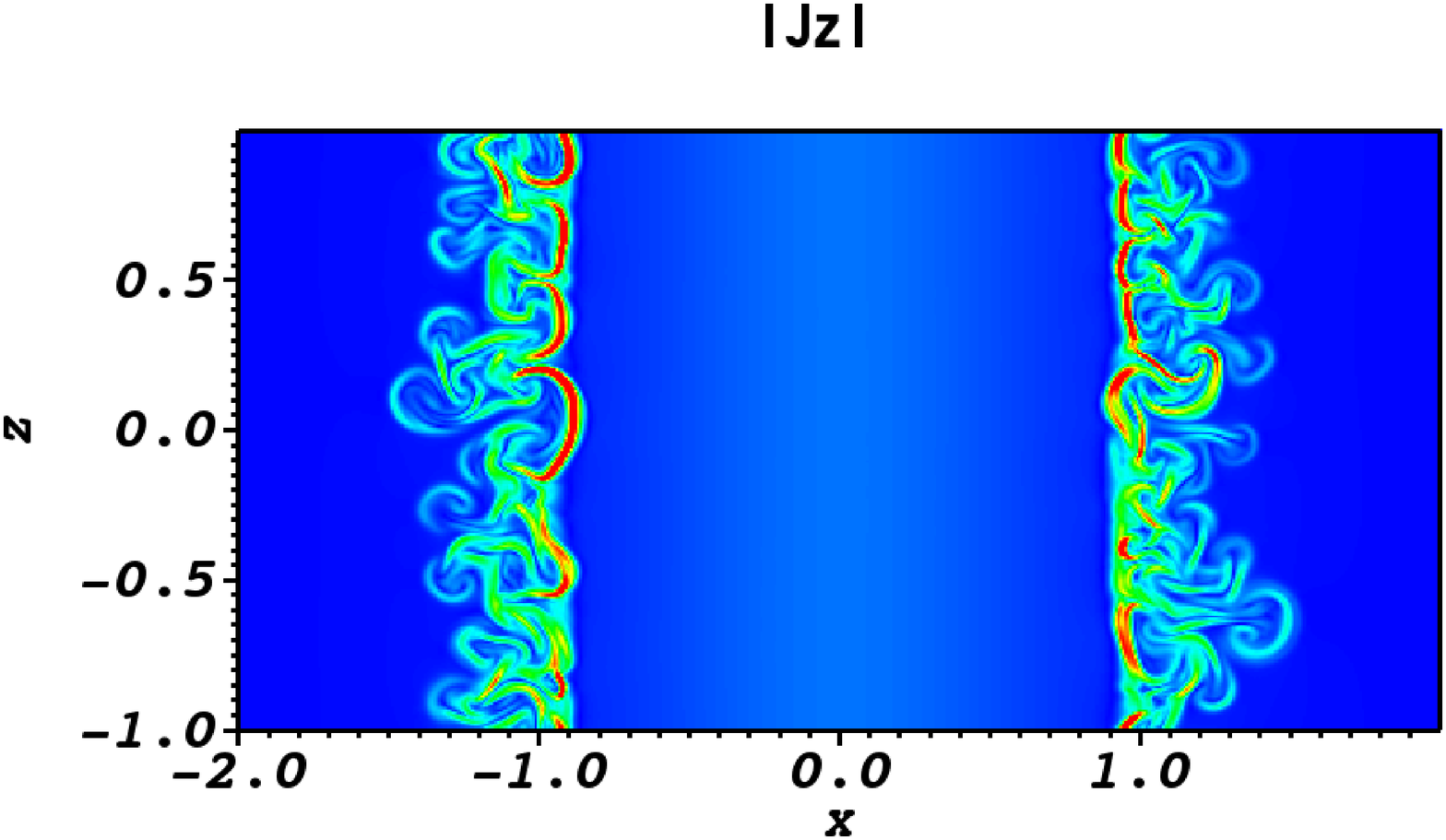}
 \quad
 \includegraphics[width=5.5cm, height = 3.45cm]{\fpath/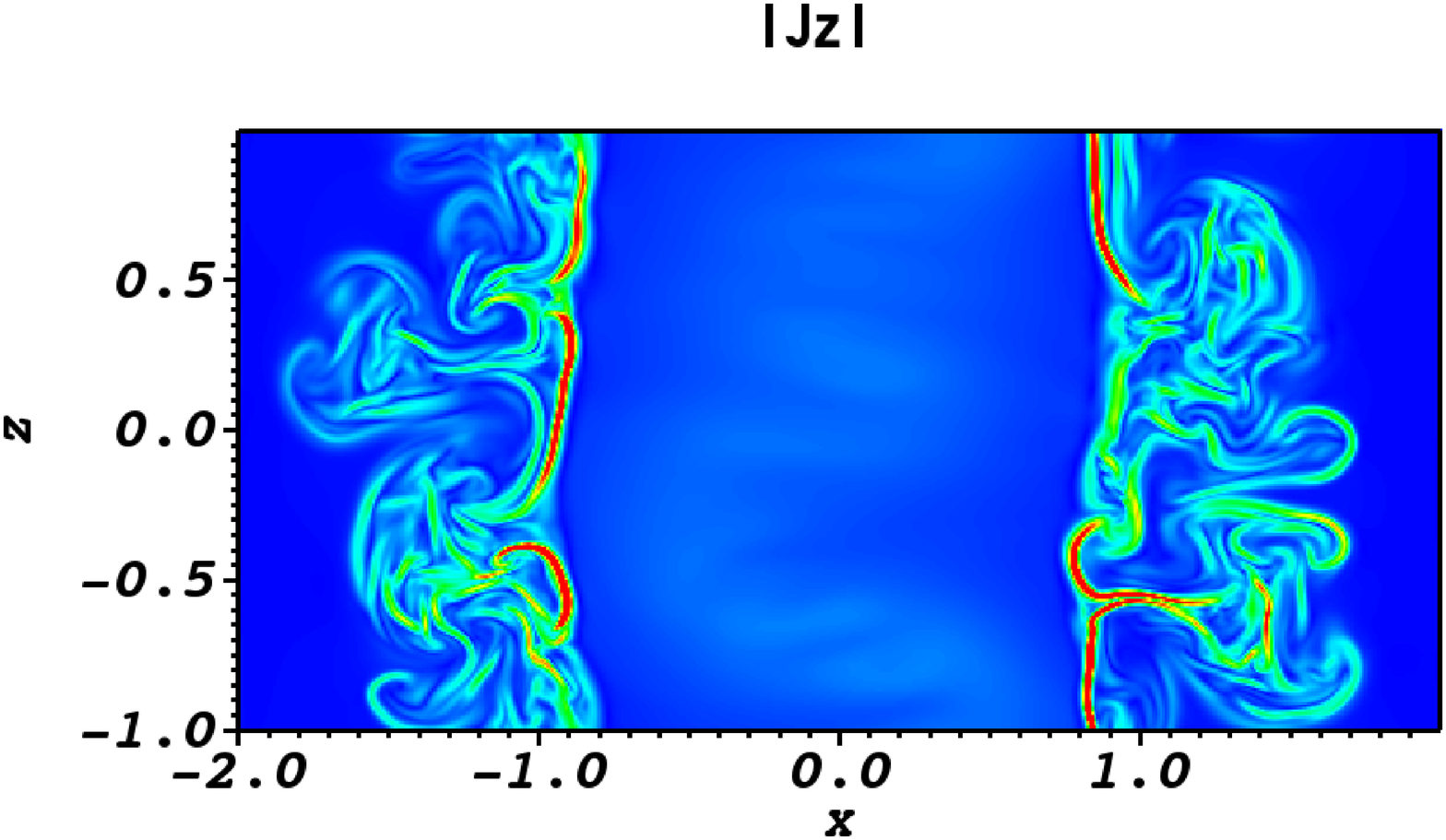}\\
 \vspace{0.5cm}
 \includegraphics[height = 5.2cm]{\fpath/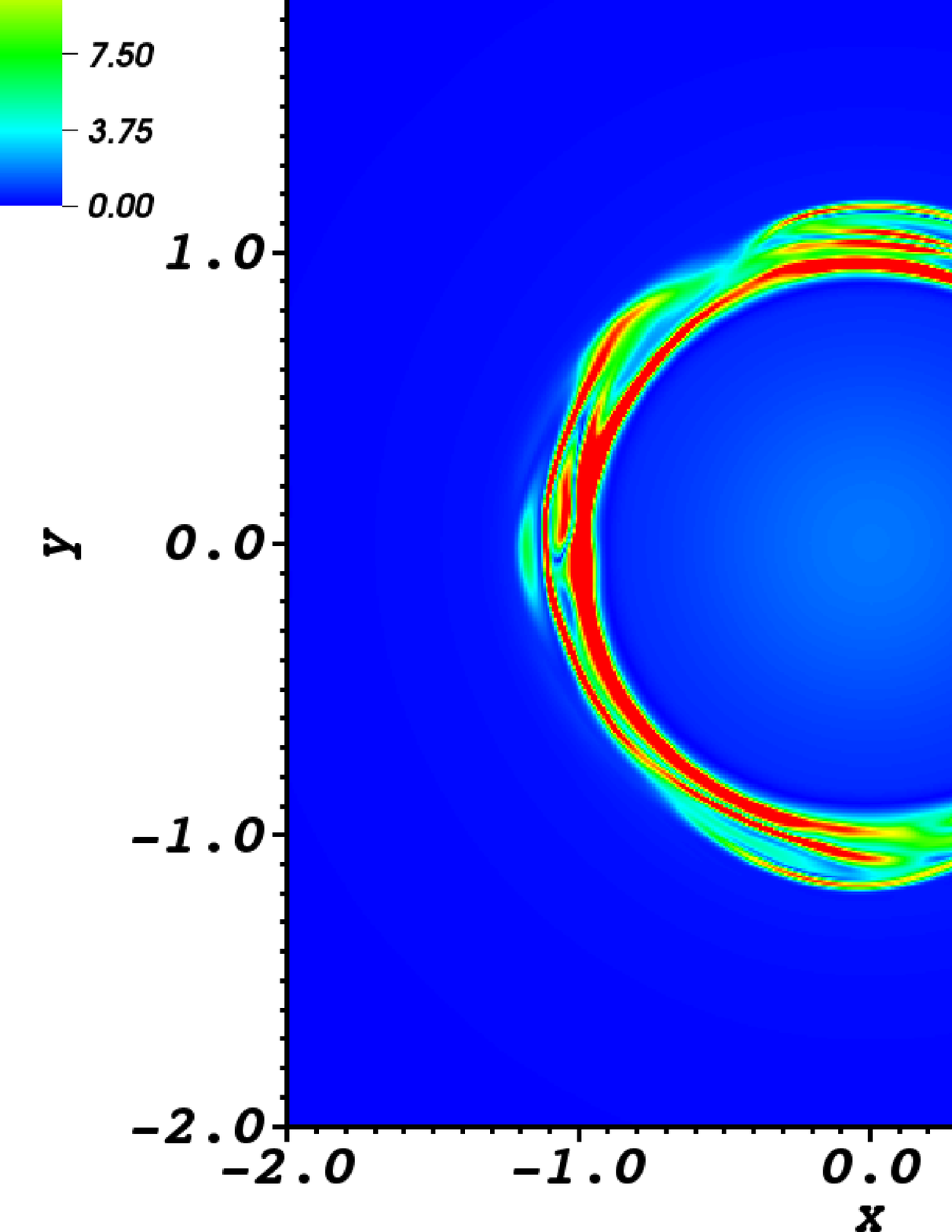}%
 \quad
 \includegraphics[height = 5.2cm]{\fpath/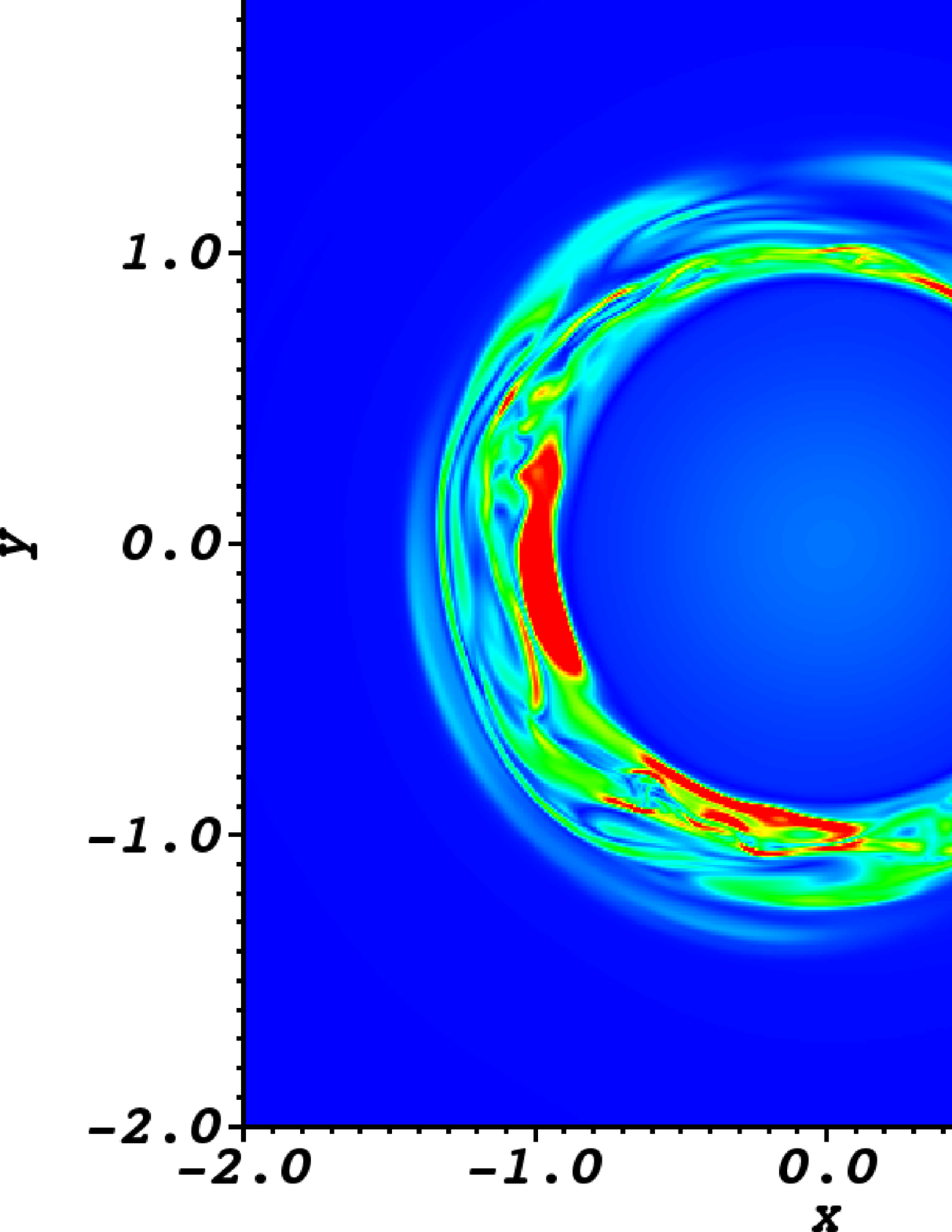}
 \quad
 \includegraphics[height = 5.2cm]{\fpath/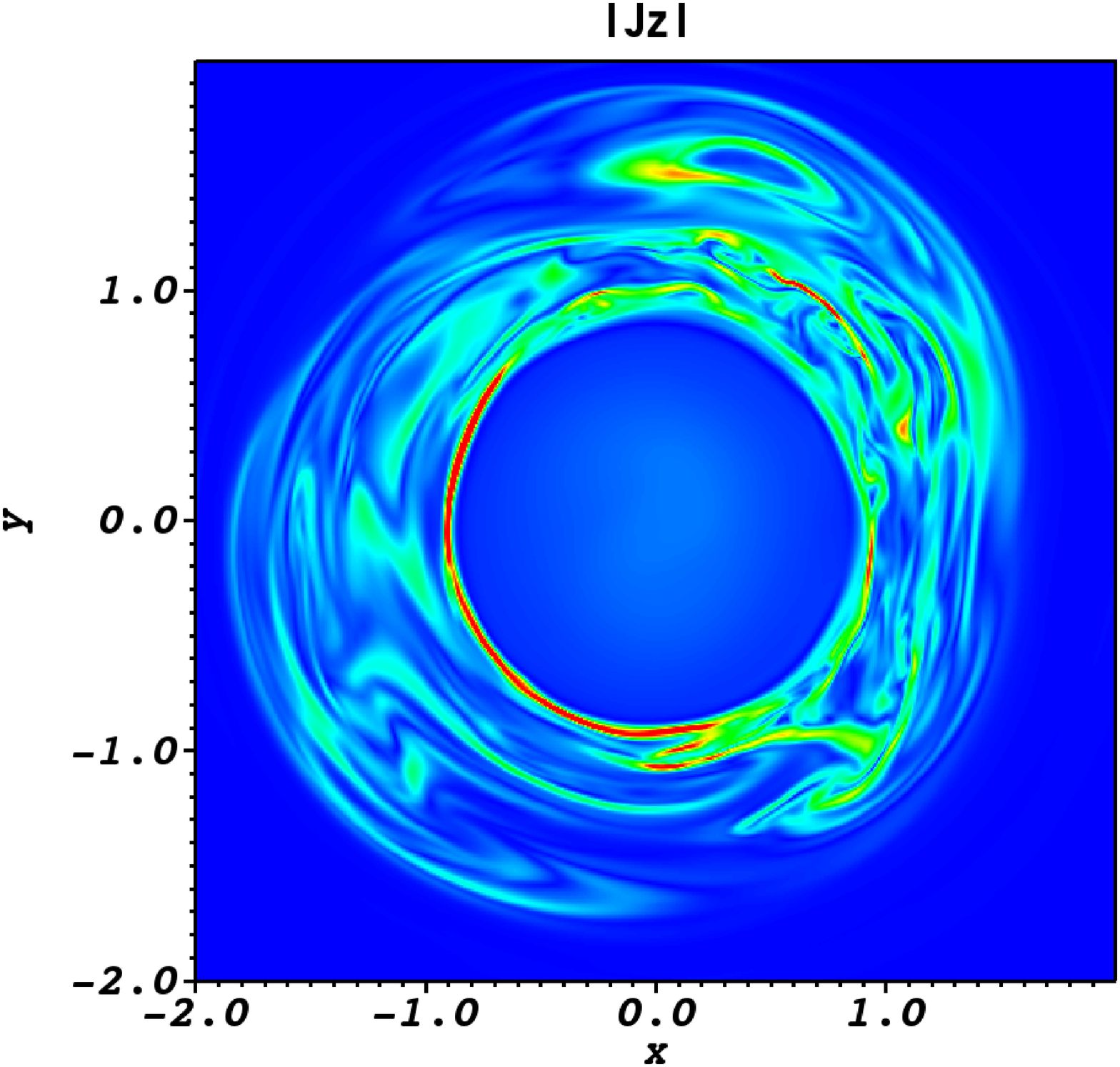}
 \caption{Snapshots for case PB-0 and $S = 2.4 \times 10^4$ for $t = 0.7 t_A$, $t = 1.0 t_A$, and $t = 1.6 t_A$.
 We show the three-slice (pseudocolor rendering) of the density \textit{(top)}, the two-dimensional slice on the $yz$  plane (at $x = 0)$ of the current density \textit{)center)},
 and the two-dimensional slice on the $xy$ plane (at $z = 0.2$) \textit{(bottom)}}\label{Fig:3DPB}
\end{figure*}

\subsubsection{Standard Harris Sheet}
\label{sec:harris_sheet}
%
%
%

According to the Sweet-Parker theory, the reconnection rate $\eta$ can be written as
\begin{equation}
\eta \equiv \frac{u_{in}}{u_{out}} \sim \frac{\delta}{L} \frac{1}{\sqrt{S}}
\end{equation}
where $u_{in}$ and $u_{out}$ are the inflow and outflow speeds, while
$\delta$ and $L$ are the current sheet's half width and half length, respectively.
\ess{To compute $\delta$ we estimate the peak value of the current density at the reconnecting region, located at $y = 0$. 
We then define $\delta$ as the distance (along $y$) where $J$ decreases by a factor $1/e$ of its peak value ($e$-folding distance), similarly to \cite{Mignone2012}.}
In Fig. \ref{Fig:Cartesian} we plot $\delta$ for various $S$ (blue circles) along
with a best fit. The Sweet-Parker scaling, $\sim S^{-1/2}$, is plotted with the black dashed line.
For comparison, we calculated the reconnection rate by estimating \ess{the rate at which the magnetic energy dissipates, 
in a manner similar to \cite{Gordovskyy2010b} and \cite{Oishi2015}. 
In order to do so, we plot the temporal evolution of the total magnetic energy in the domain,}
and we compute the slope $\gamma= dE_m/dt$, \ess{where $E_m$ is normalized to the initial value of the magnetic energy and $t$ to the Alfv\'en time.}
The slopes are calculated at $t = t_{A}$ so as
to ensure that magnetic reconnection has already started \footnote{We have checked that the choice of the time at which slopes 
are computed has hardly noticeable differences on the results}.
The \rev{dissipation rate}, $\gamma$, for different values of $S$ is shown in Fig. \ref{Fig:Cartesian} (green circles).
The two estimates of the reconnection rate
are compatible and in agreement with the Sweet-Parker scaling.
\rev{For convenience, in this work we will measure the dissipation rate $\gamma$, that is equivalent to the reconnection rate only for cases where reconnection is 
the dominant dissipating process. We will discuss the implications of this choice in the last paragraph.}

\subsubsection{Circular Current Sheet}\label{Par:Polar2D}
%
%
%

The simulations exhibit different features for increasing values of the Lundquist number, \rev{estimated using a characteristic length $L = 2\pi r_1$ }.
For $S = 1.7 \times 10^3$ the layer is \rev{highly} diffusive, and there is no evidence of island formation.
For larger $S$ magnetic reconnection results in the formation of four plasmoids that further move along the curved
current sheet and later merge into a larger island.

\begin{figure*}
 \centering
 \includegraphics[width=8.cm]{\fpath/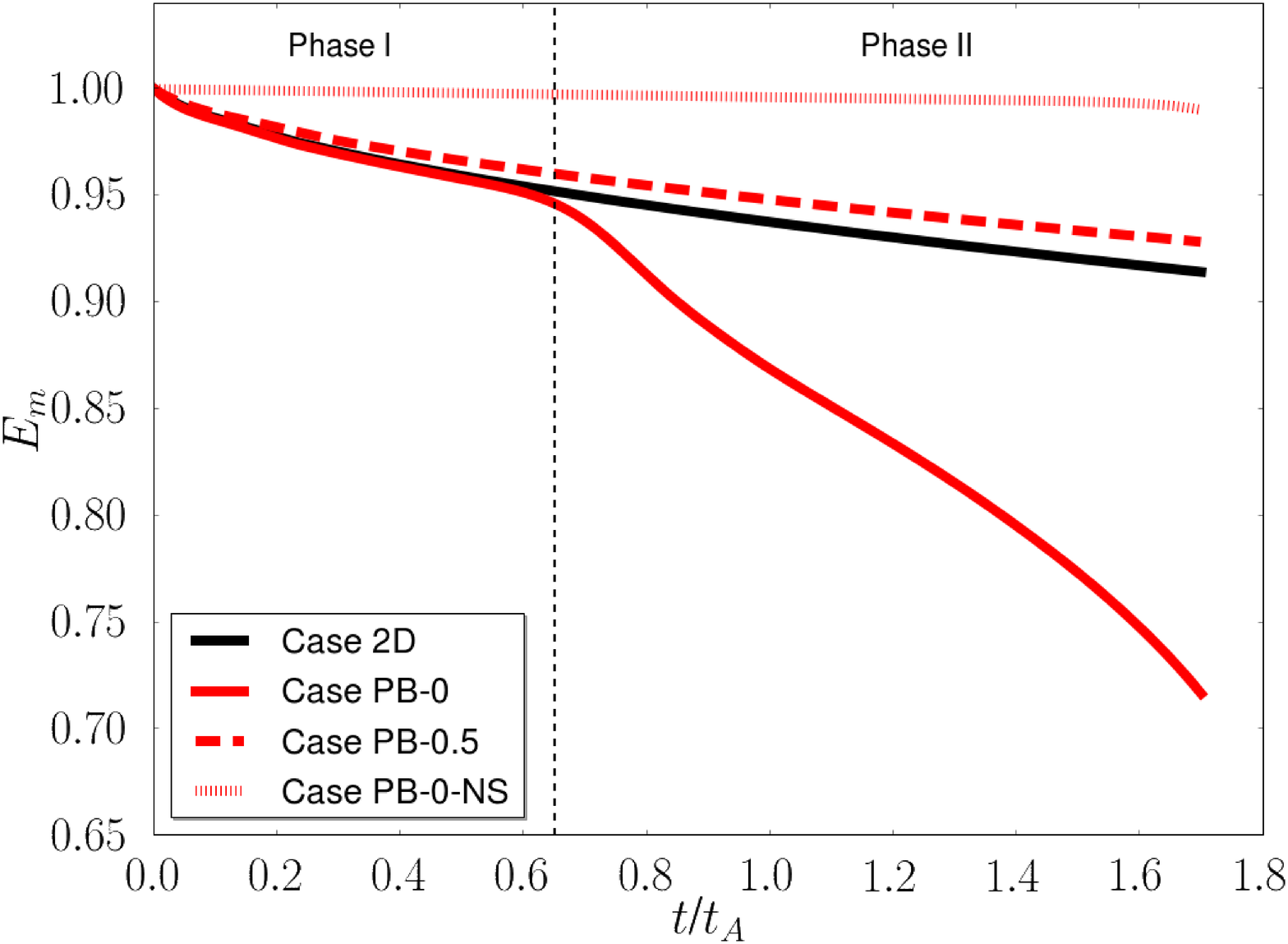}
 \includegraphics[width=8.5cm]{\fpath/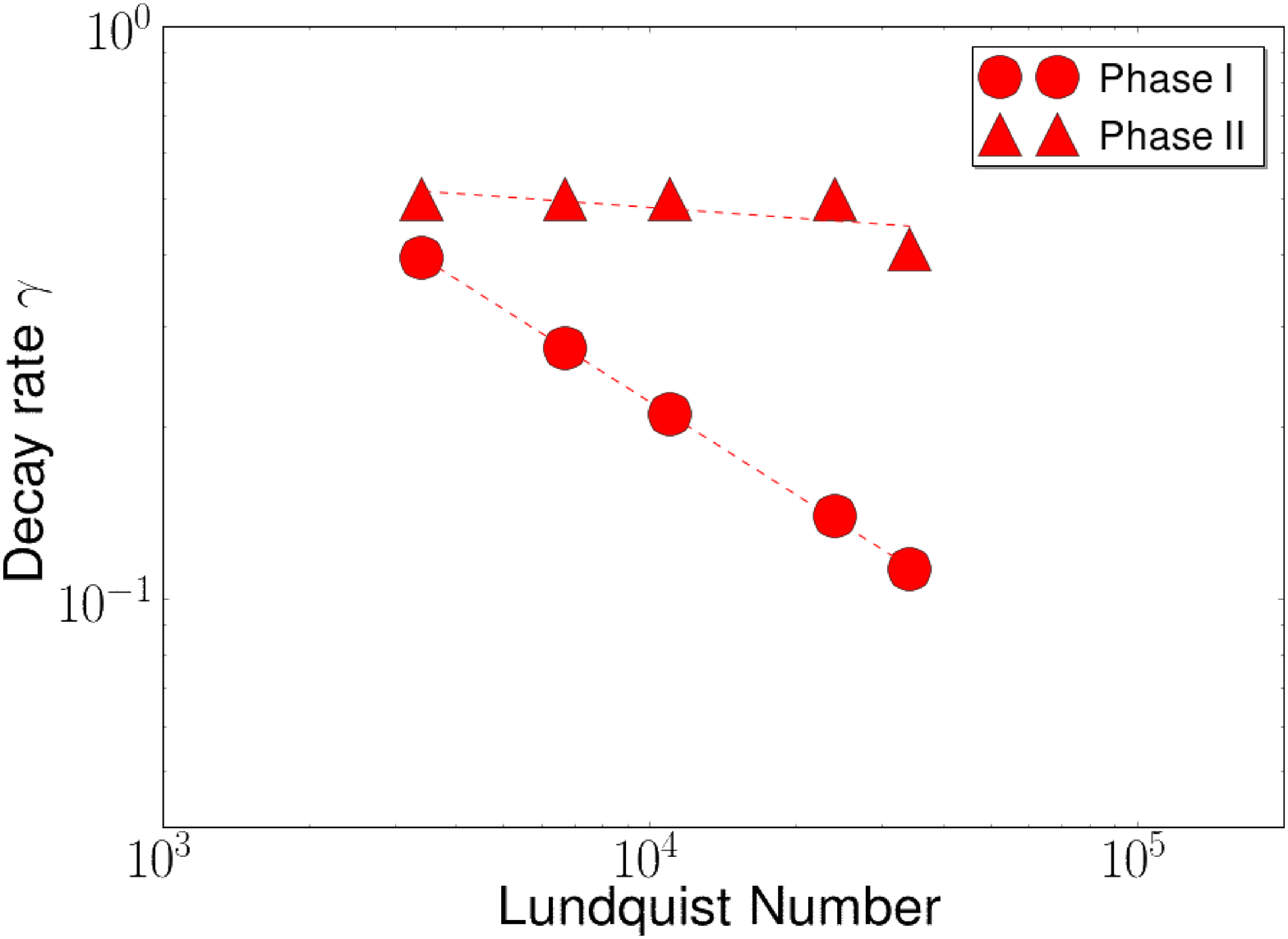}
 \caption{\textit{(Left)} Temporal evolution of the volume averaged magnetic energy normalized to the initial magnetic energy for 
 case PB-0 (red solid line), case PB-0.5 (dashed line) and case PB-0-NS (dotted line), along with case 2D (black).
 For all cases the Lundquist number is $S = 2.4 \times 10^4$.
 The black vertical line separates phase I and phase II for case PB-2D.
 \textit{(Right)} Decay rate $\gamma$ of the magnetic energy as a function of $S$
 for case PB-0. For each $S$ value,
 $\gamma$ is computed in \es{phase I} (circles) and \es{phase \rom 2} (triangles).
 We note that the decay rate follows the Sweet-Parker scaling during phase I, and is nearly independent on $S$ in phase II.}\label{Fig:DissipationPB}
 \end{figure*}

When $S > S_c \simeq 1 \-- 5 \times 10^4$ the simulations show evidences of multiple fragmentation
of the current sheet by secondary tearing instability. Here we see the continuous hierarchical formation
of islands between two already formed larger islands (see Fig. \ref{Fig:2D}).
The islands further move away from the region of creation and feed the growth of a monster island \rev{\cite[see][]{Uzdensky2010}}.
The fragmentation of the layer ultimately results in formation of several small sized current sheets randomly oriented (see Fig \ref{Fig:2D}, right side) \rev{leading the system into a ``plasmoid turbulence'' phase, 
as described in \cite{Loureiro2012}. At this stage the current sheet configuration resembles that of the simulations of \cite{Kowal2009} and \cite{Loureiro2009} 
\footnote{\rev{We note, however, that these models focused on magnetic reconnection in the presence of a pre-existing, background turbulence.}}.}

In Fig. \ref{Fig:2D_Em_J2} (upper panel) we show the temporal evolution of the average magnetic energy normalized to the initial magnetic energy for
different values of the Lundquist number.
For cases with $ S \ga 10^4$ the decay rate sharply increases at $t \simeq 3.2 t_A$ (seen as steepening of the slope). This corresponds
to the time when the secondary tearing instability sets in.
\ess{Consequently we} define two different phases in the temporal evolution of the magnetic energy: phase I that starts at the beginning of the simulation
until the onset of the (plasmoid) instability, and phase II that starts after the onset of the instability
(note that phase II is present only in the case where $S \geq 10^4$).

The \rev{central} panel of Fig. \ref{Fig:2D_Em_J2} shows the temporal evolution of the average value of ${J^2}$ (denoted by $\av{J^2}$), where $J$ is the current density.
There is a clear distinction in the growth of $\av{J^2}$ with time between the two phases mentioned above, especially
for higher values of $S$. The evolution of $\av{J^2}$ in phase I, \ess{after an initial transient,} is typically flat for
all values of $S$. As the instability sets in, a sharp rise is seen in $\av{J^2}$ for values of $S \geq 10^4$, while
for smaller $S$ values it continues to remain flat. In particular for $S = 1\times 10^5$ (yellow curve), $\av{J^2}$ increases steeply by a factor of four after $t = 3.2 t_A$.

\rev{Finally, the inset panel of Fig. \ref{Fig:2D_Em_J2} shows the temporal evolution of the Ohmic heating. We note that while in phase I $\eta \av{J^2}$ 
decreases for larger values of $S$, in phase II it becomes independent on the Lundquist number.}

\ess{By} fitting the curves representing the temporal evolution of $E_m$ during phase I and phase II, we can estimate
the decay rate $\gamma$ for the two phases.
We show in Fig. \ref{Fig:Reconnection_Rate_Polar} the decay rate for different values of $S$.
We see that the \rev{dissipation rate} follows the Sweet-Parker scaling
\ess{during} phase I (circles), and it is nearly independent of $S$ in phase II (triangles).
One can therefore identify phase I with a Sweet-Parker phase and phase II with a ``fast reconnection'' \ess{regime}.
\ess{The rate at which magnetic energy is dissipated during the fast reconnection regime is \rev{$\sim 0.1 t_A^{-1}$}, consistent with rates reported 
by previous numerical results \citep[e.g.,][]{Huang2010,Bhattacharjee2009,Loureiro2016} and the theoretical model of \cite{Uzdensky2010}.}
We emphasize that this ``fast reconnection'' regime is related to the plasmoid instability, that sets in only when $S \geq 1 \times 10^4$.
These results confirm the findings of many earlier papers studying magnetic reconnection in a Harris current sheet
for high $S$ values \citep[see, e.g.,][]{Huang2010,Huang2013}, and extend them to the case of a circular current sheet.

\subsection{Three-Dimensional Cases}
%
%
%
%

\subsubsection{3D pressure-balanced}\label{sec:PB}
%
%
%
%

\begin{figure*}
 \centering
 \hspace{-0.3cm}\includegraphics[height = 0.63\columnwidth]{\fpath/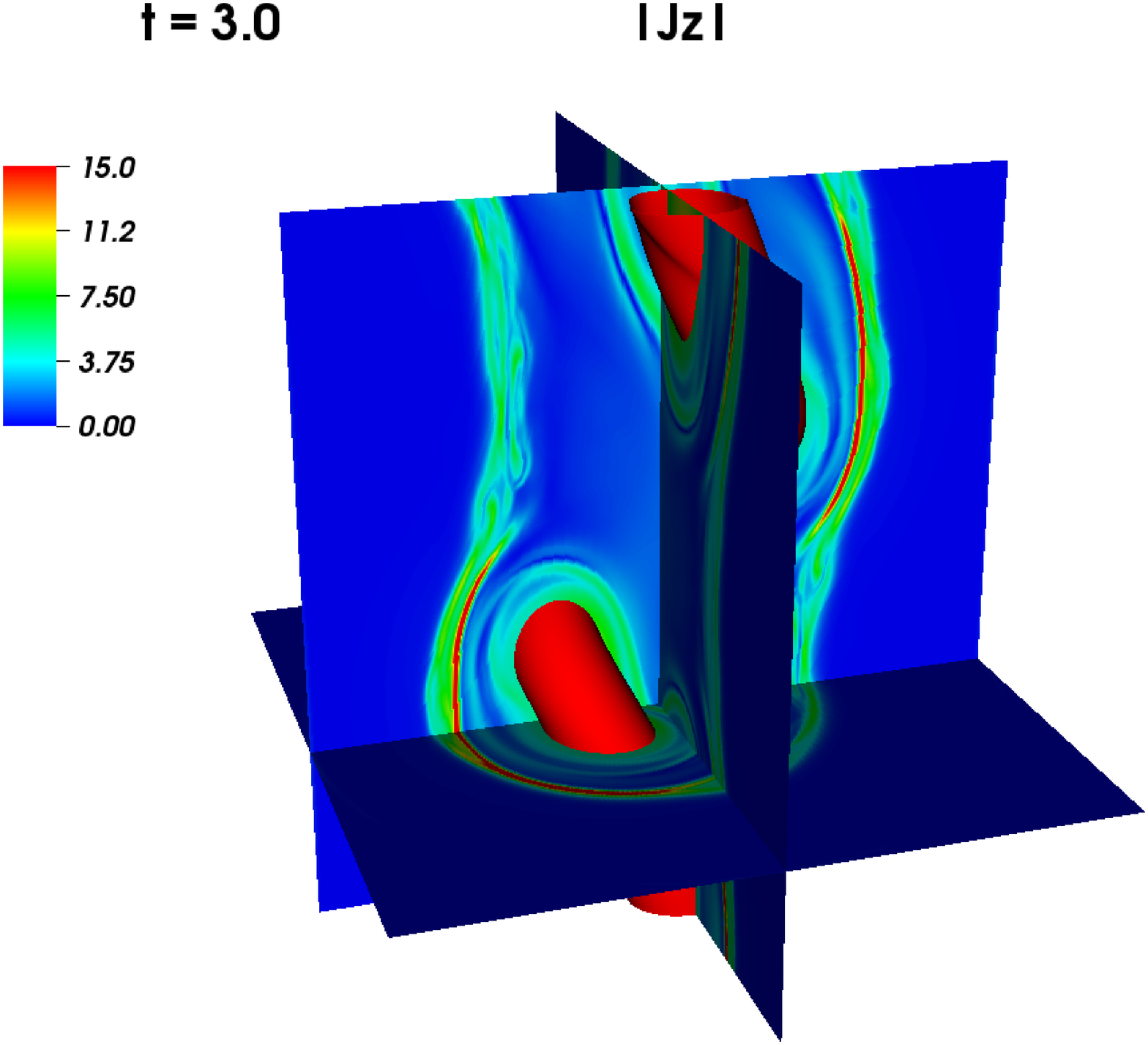}%
 \quad\quad
 \includegraphics[height = 0.63\columnwidth]{\fpath/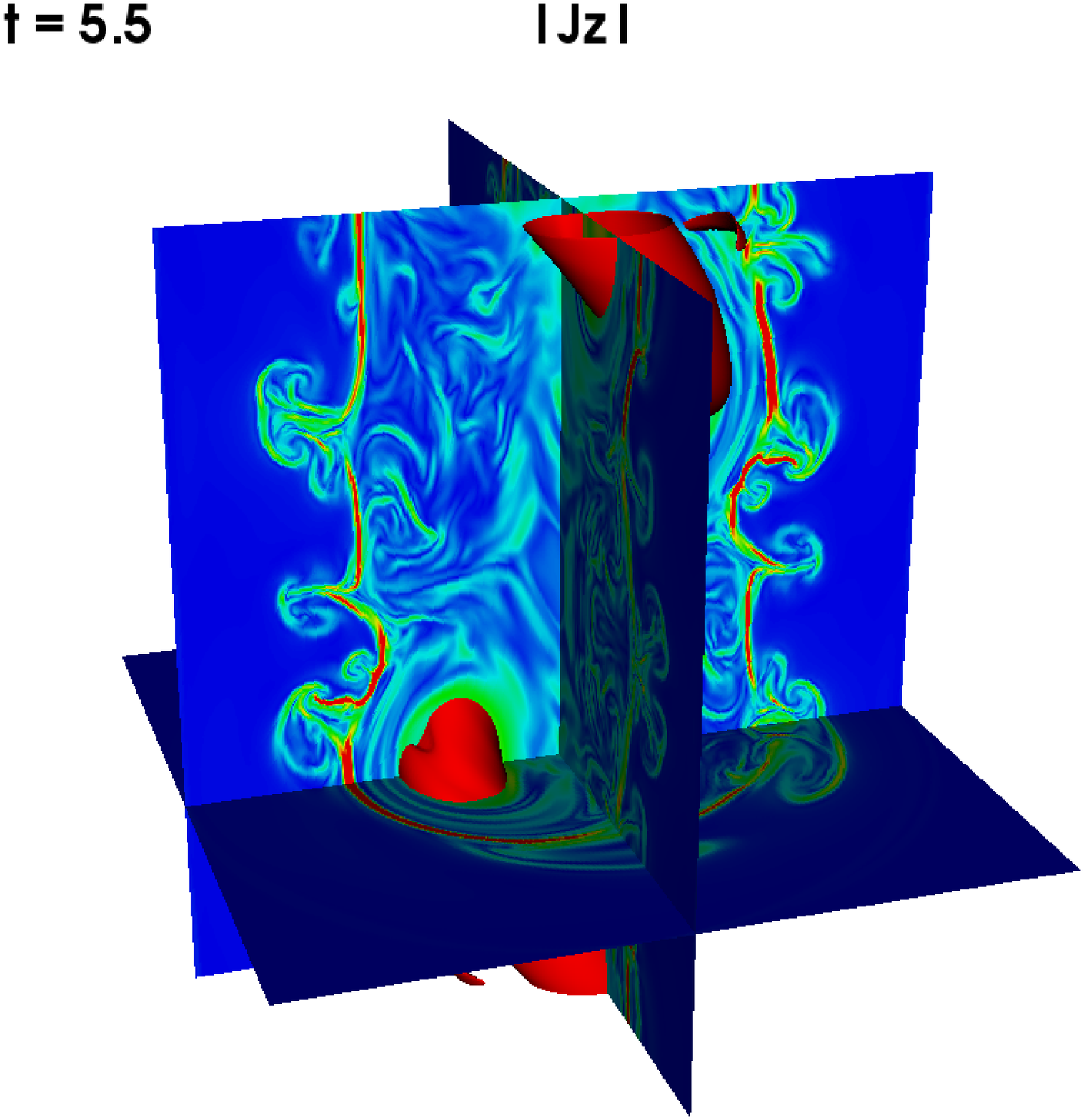}
 \quad
 \includegraphics[height = 0.63\columnwidth]{\fpath/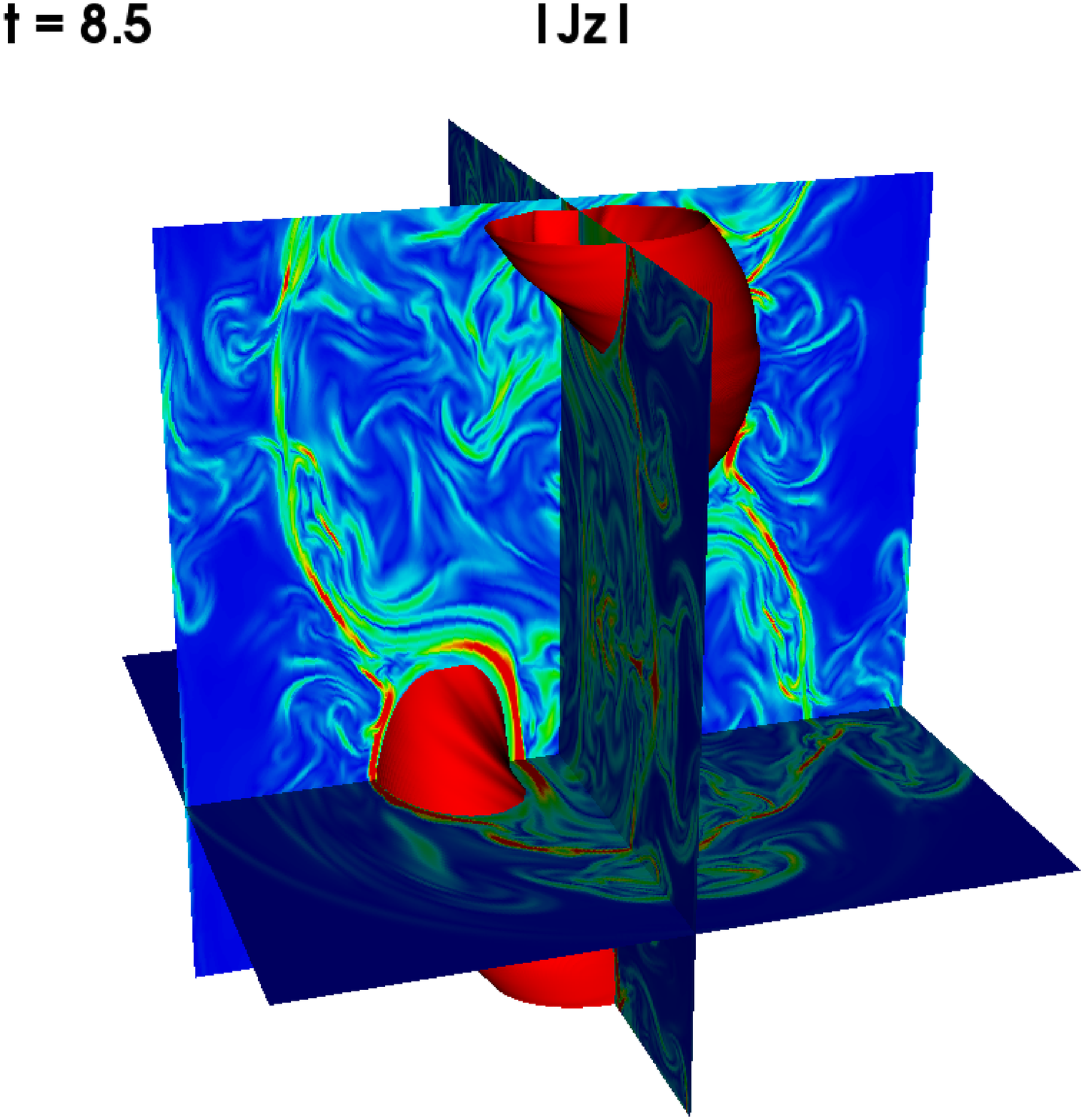}\\
 \vspace{0.5cm}
 \includegraphics[height = 0.62\columnwidth]{\fpath/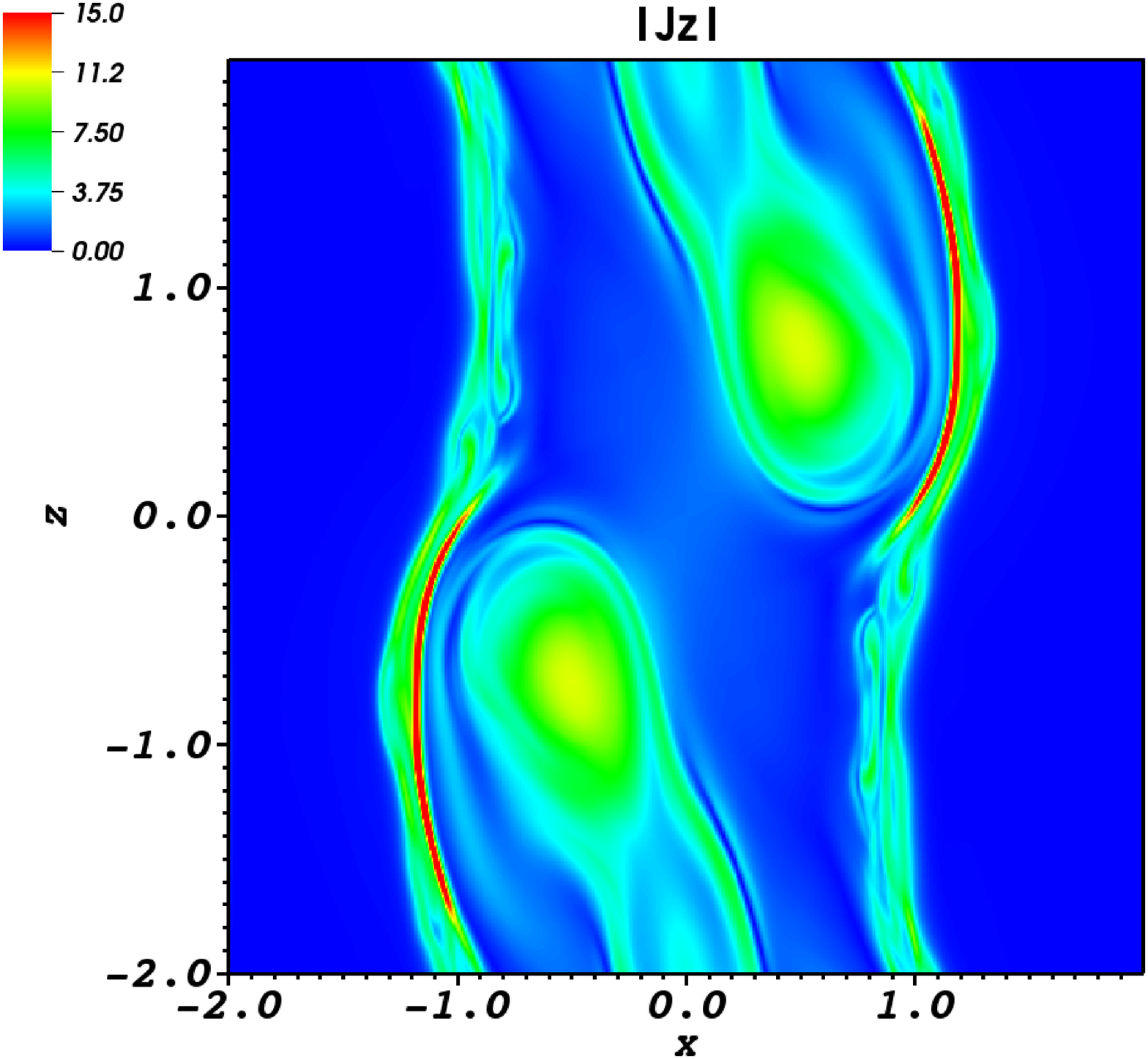}%
 \quad
 \includegraphics[height = 0.62\columnwidth]{\fpath/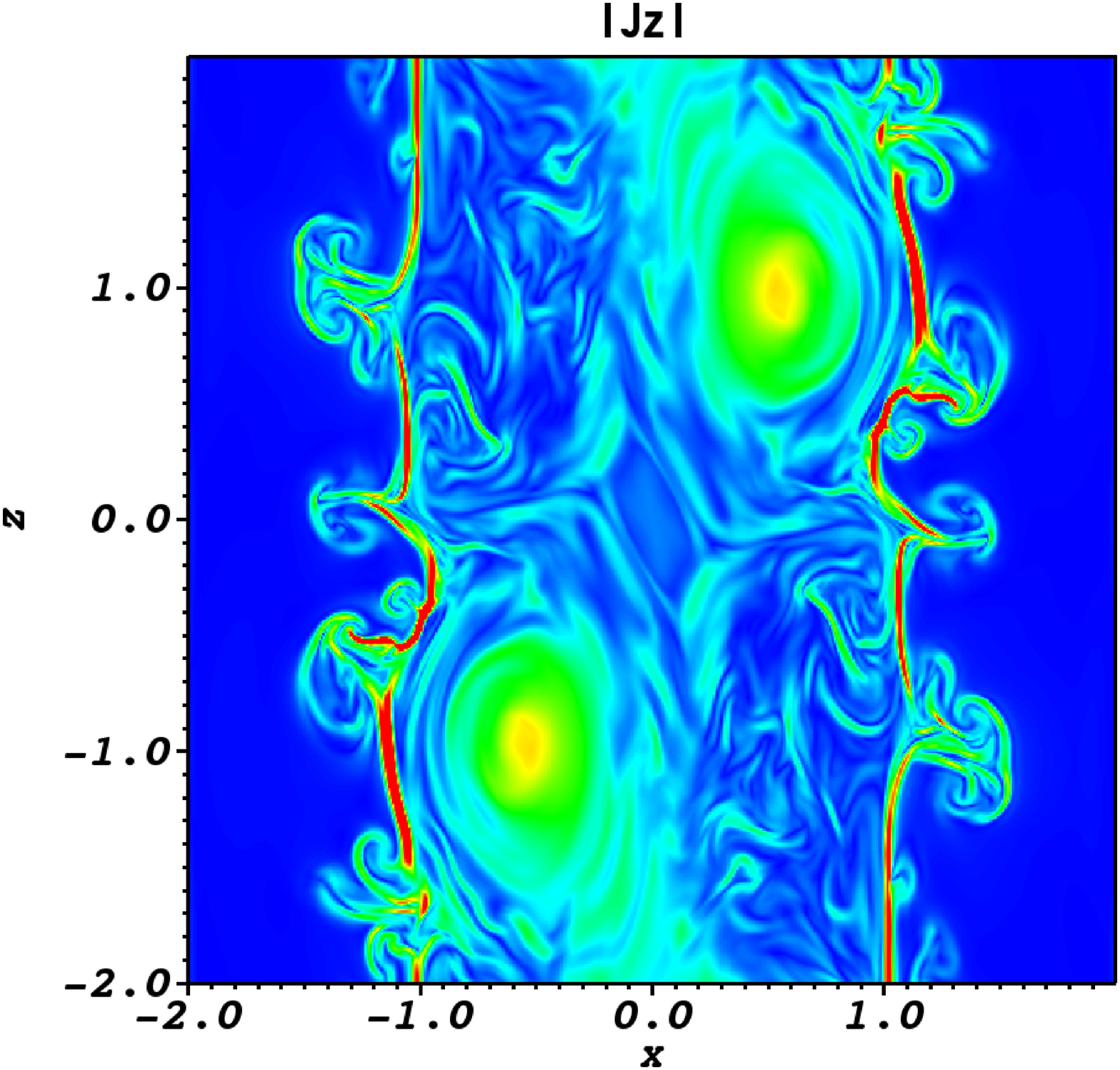}
 \quad
 \includegraphics[height = 0.62\columnwidth]{\fpath/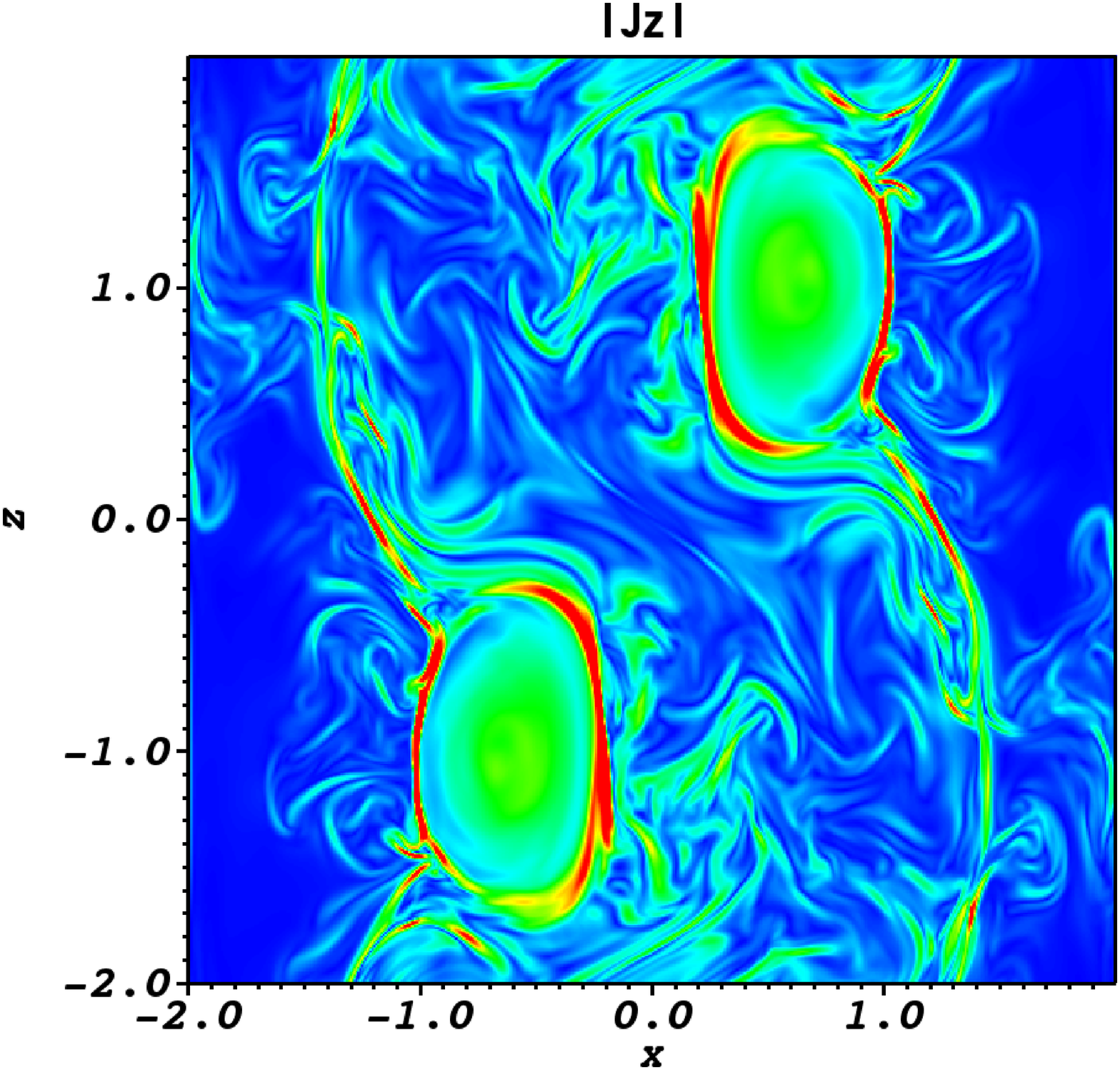}\\
 \vspace{0.5cm}
 \includegraphics[height = 0.62\columnwidth]{\fpath/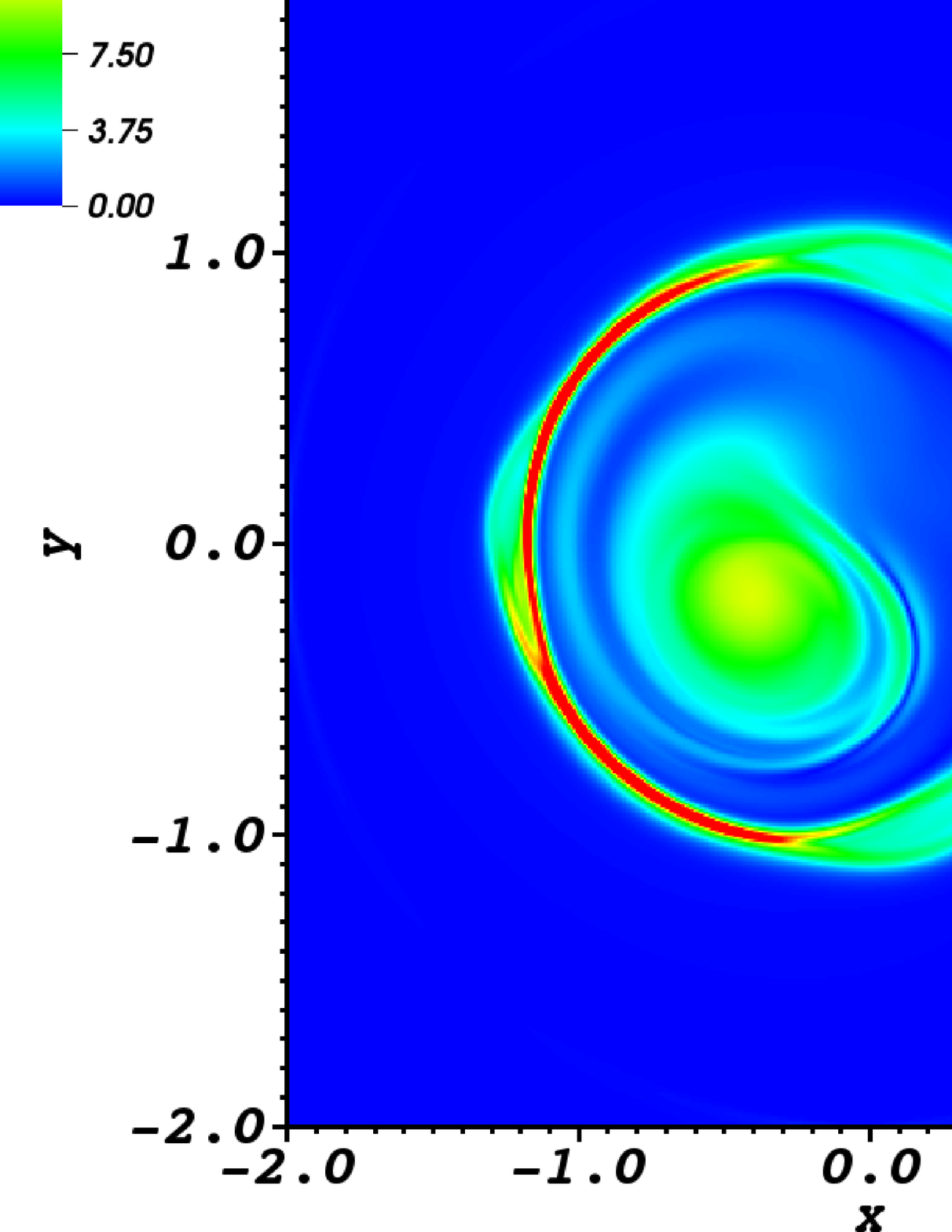}%
 \quad
 \includegraphics[height = 0.62\columnwidth]{\fpath/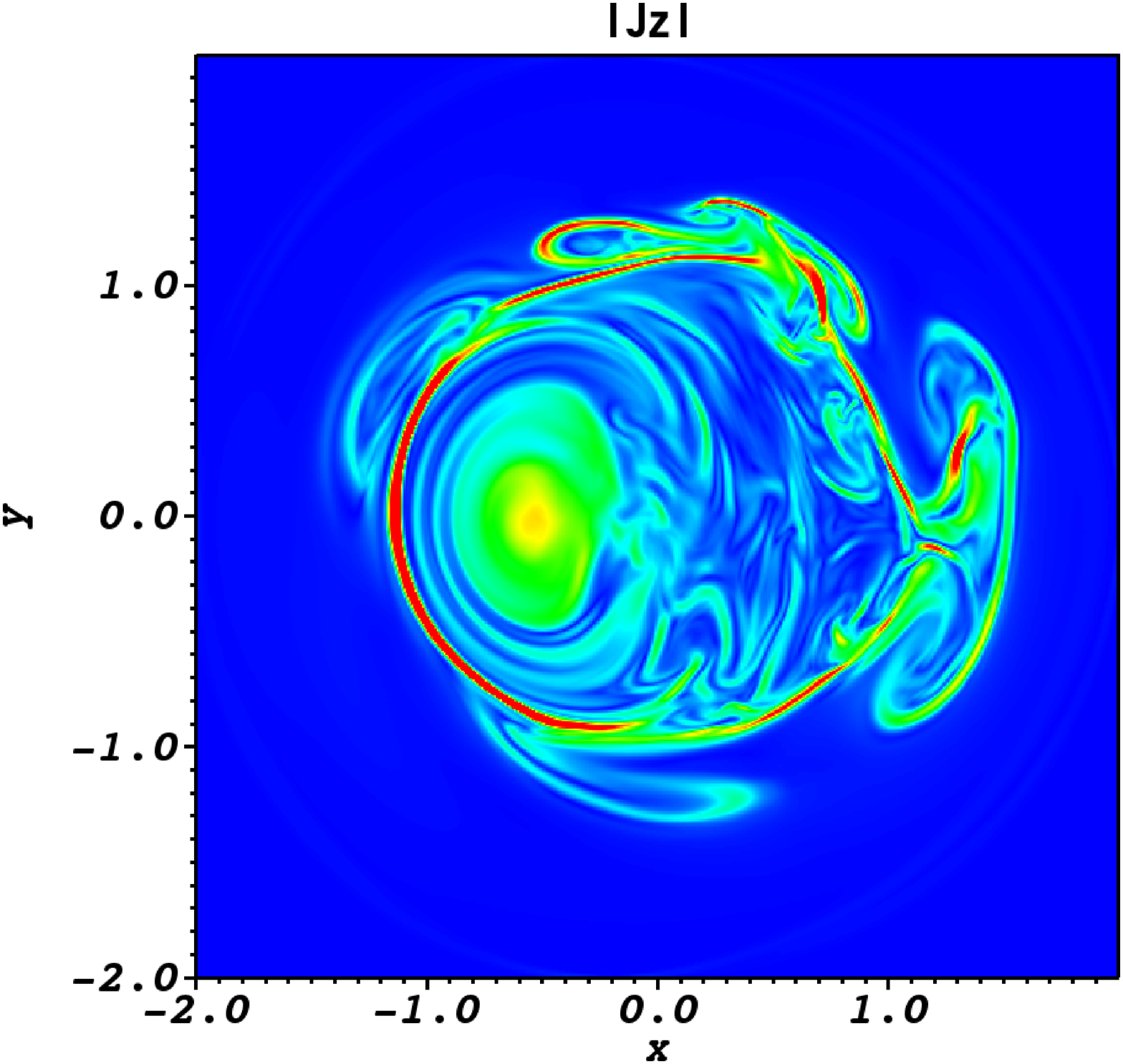}
 \quad
 \includegraphics[height = 0.62\columnwidth]{\fpath/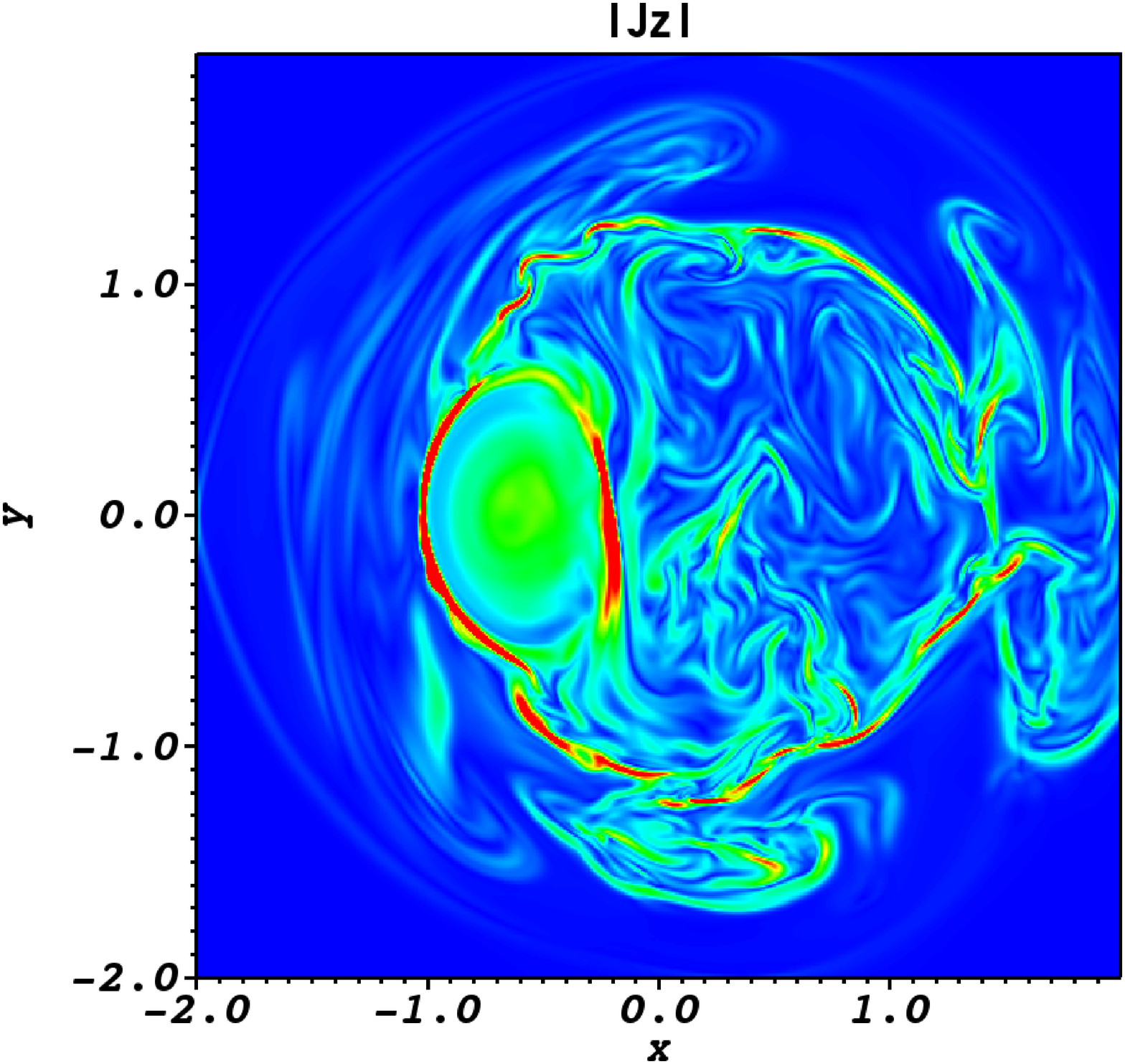}
 \caption{Snapshots for case FF-0.2 with $S = 2.4 \times 10^4$ at $t = 6 t_A$, $t = 11 t_A$ and $t = 17 t_A$. 
  We show the three-slice rendering of the current density superimposed to the contour rendering of the density \textit{(top)},
  the two-dimensional slices on the $xz$ plane (at $x = 0$) of the current density \textit{(center)},
  and the two-dimensional slices on the $xy$ plane (at $z = -1$) of the current density \textit{(bottom)}. 
  }\label{Fig:3DFF}
\end{figure*}

The PB cases are unstable to PDI
that triggers the formation of finger-like structures at the current sheet.
The three-slice rendering of the density
for case PB-0 with $S = 2.4\times 10^4$ at $t = 0.7 t_A$ (left), $t = 1.0 t_A$ (center) and
$t = 1.6t_A$ (right) is shown in the top panels of Fig. \ref{Fig:3DPB}, where one can observe the formation and growth of the fingers.
These features start to form at $t \simeq 0.6t_A$
and continue to develop during the simulation.
The growth of the PDI leads to the total disruption of the plasma column
for $t > 3.5 t_A$.
The central panels of Fig. \ref{Fig:3DPB} show the 2D slice of the current density
on the $yz$ plane (at $x = 0$).
Peaks of the current density in correspondence of the fingers can be noticed, pointing out that each of these features may become secondary current sheets
where magnetic reconnection \ess{takes place.}
The 2D slice of the current density on the $xy$ plane (at $z = 0.2$) is shown in the lower panels of Fig. \ref{Fig:3DPB}.
The formation of the fingers results in the fragmentation of the current sheet in the $xy$ plane in several secondary and small-sized current sheets
with high values of the current density. The length $L'$ of these secondary current sheets is much smaller than
the size $L$ of the original current sheet, therefore yielding \ess{an effective} value of the Lundquist number $S' \propto L' \ll S$ and, finally, a
\rev{dissipation rate} that does not depend on $S$.
This can be seen on the left side of Fig. \ref{Fig:DissipationPB} where the temporal evolution of the volume averaged magnetic energy (normalized to its initial value) 
for case PB-0 (red solid line), case PB-0.5 (red dashed line) and case PB-0-NS (red dotted line), along with case PB-2D (black solid line), is shown.
The Lundquist number is $S = 2.4 \times 10^4$ for each of these plots.
At early times the dissipation of the magnetic energy for case PB-0 overlaps with the 2D case.
The curve is then characterized by a sharp change of the slope at $t \simeq 0.65 t_A$, i.e. the time at which the finger-like features of the PDI start to form.
In analogy with the PB-0-2D case, we define two different \es{phases} in the temporal evolution of the magnetic energy:
\es{phase I} that starts at the beginning of the simulation until formation of the features of the pressure-driven instability,
and \es{phase II} that begins after the formation of such features.
In order to study the scaling of the \rev{dissipation rate} for this case, we computed the decay rate $\gamma = dE_m/dt$ (see sec \S\ref{sec:harris_sheet})
by estimating the slope of the magnetic dissipation both in phase I and in phase II.
Figure \ref{Fig:DissipationPB} (right panel) shows the decay rate $\gamma$ for different values of $S$
and for the two different phases. In analogy with the 2D case, the \rev{dissipation rate} follows the Sweet-Parker scaling
in phase I (circles), and it is nearly independent on $S$ (a linear fit yields a slope $\simeq 0.06$) in phase II (triangles).
\ess{The rate of dissipation of magnetic energy during phase II is \rev{$\sim 0.5 t_A^{-1}$}, somewhat larger than the rate obtained from 2D simulations (see section \S \ref{Par:Polar2D}).}
\begin{figure*}
 \centering
 \includegraphics[width=8.2cm]{\fpath/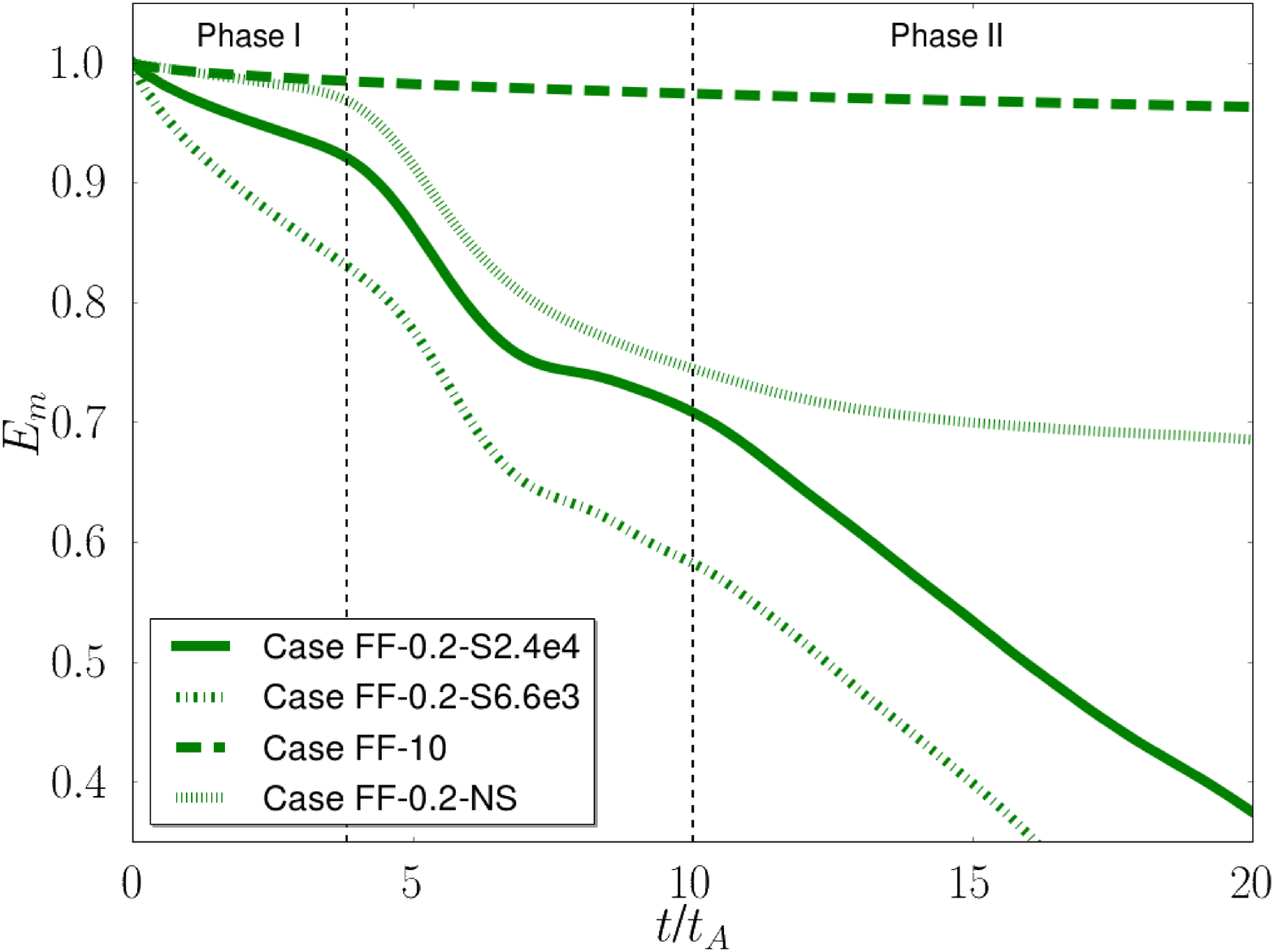}
  \includegraphics[width=8.9cm]{\fpath/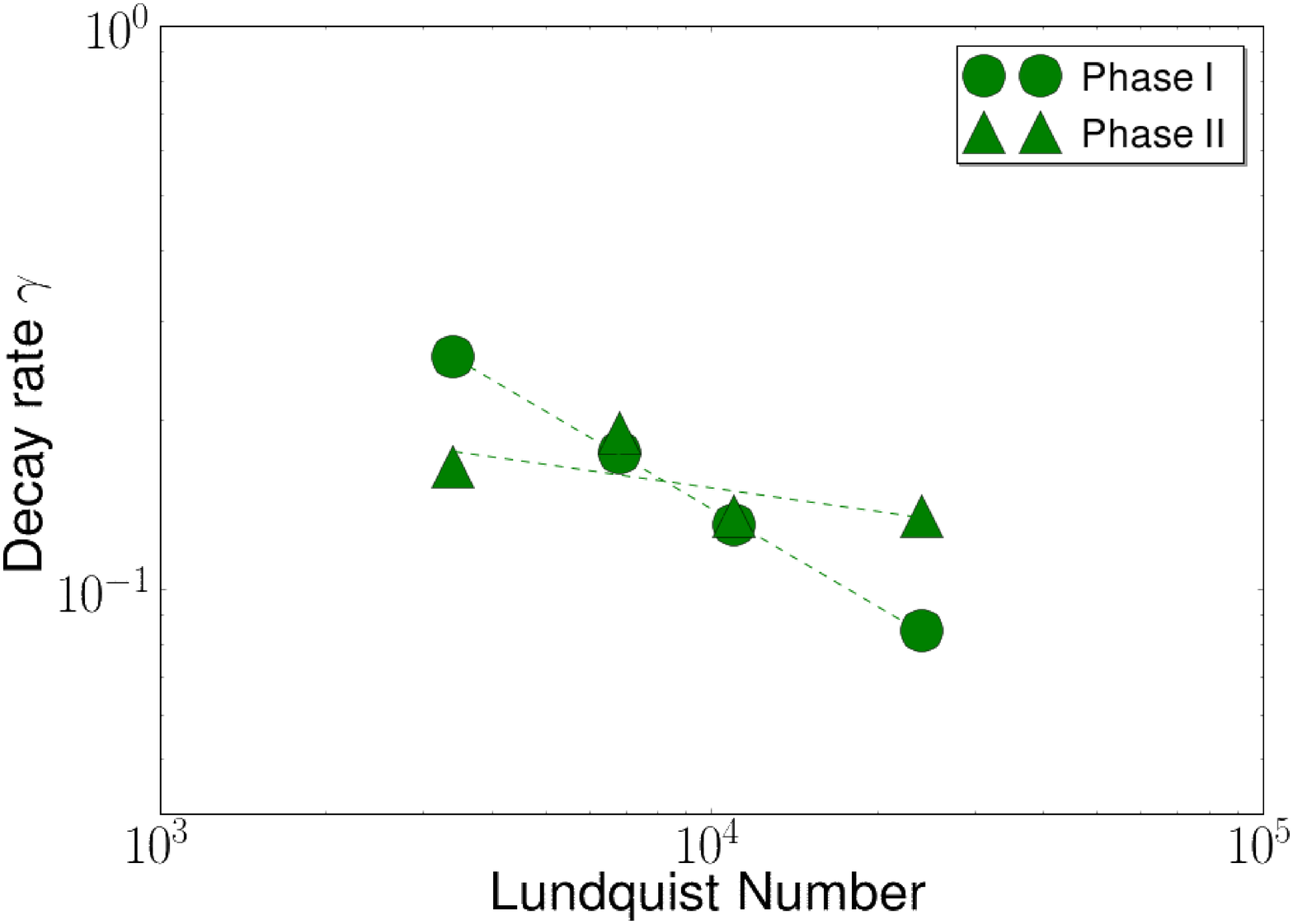}
 \caption{\textit{(Left)} Temporal evolution of the volume averaged magnetic energy normalized to the initial magnetic energy for
 case FF-02 with $S = 2.4 \times 10^4$ (solid line),
 case FF-0.2 with $S = 6.6 \times 10^3$ (dash-dotted line), case FF-10 with $S = 2.4\times 10^4$ (dashed line) and case FF-0.2-NS with $S = 2.4\times 10^4$ (dotted line).
 The black vertical dashed lines indicates the beginning of phase II, where the decay rate $\gamma$ for this
 case was computed.
 \textit{(Right)} Decay rate $\gamma$ of the magnetic energy as a function of $S$
 for case FF-0.2. For each case $\gamma$ is computed in \es{phase I} (circles) and \es{phase \rom 2} (triangles).
 We note that the decay rate follows the Sweet-Parker scaling during phase I, and is
 nearly independent on $S$ in phase II.}\label{Fig:DissipationFF}
 \end{figure*}
The temporal evolution of the PDI for case PB-0.5 is much slower, as the \ess{increased} pitch has a stabilizing effect on PDI.
Consequently, this case does not show any features related to 3D instabilities throughout the simulation,
and hints of finger formation are evident only at the very end of the simulation. The plasma column can be therefore considered
as a replication along the $z$ direction of the 2D configuration of \S \ref{Par:Polar2D}.
This can be seen in the top panel of Fig. \ref{Fig:DissipationPB}, where the dissipation of the (volume averaged) magnetic energy in the plasma column for this case
(dashed line) overlaps with the 2D case.
Finally, case PB-0-NS (dotted line) shows no sign of dissipation.
\begin{figure}
 \centering
 \includegraphics[height = 0.65\columnwidth]{\fpath/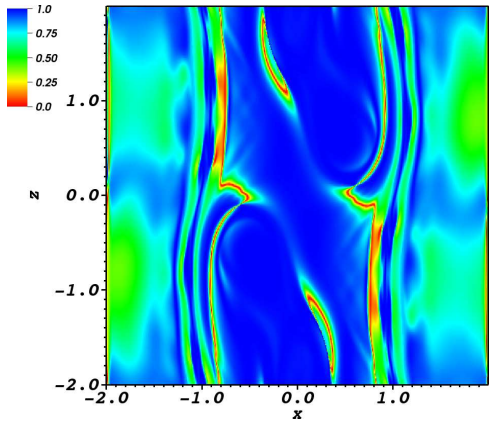}
 \caption{Two-dimensional slices on the $yz$ plane (at $x = 0$) of $\chi=$cos$^2(\theta)$ (see text) at $t = 6 t_A$ for case FF-0.2.
 Regions where $\chi = 0$ are likely unstable to PDI that can be
  responsible for the features shown in the central panels of Fig. \ref{Fig:3DFF}.}\label{Fig:3DFF_sliceY_ETA}
\end{figure}
The fragmentation of the layer due to the onset of the PDI and the consequent formation of small current-sheets observed for case PB-0 
therefore yields a \rev{dissipation rate} that does not depend on the Lundquist number.
Differently from the 2D secondary tearing instability, the onset of PDI and its features do not depend on $S$,
and the ``fast reconnection'' regime holds also for $S \simeq 1 \times 10^3 < S_c$, when the secondary
tearing instability is not present.

\subsubsection{3D force-free}
%
%
%

This configuration is unstable to CDI that favors the distortion of the
plasma column due to the growth of the kink mode.
This effect is visible in case FF-0.2, that exhibits a prominent bending
of the plasma column that begins at $t \simeq 3t_A$
and continues until the end of the simulation.
The three-slice rendering of the current density superimposed with the contour of the density
at $t = 6 t_A$, $t = 11 t_A$ and $t = 17 t_A$ for case FF-0.2 with $S = 2.4 \times 10^4$ is shown in the top panel of Fig \ref{Fig:3DFF}.

Along with the distortion of the plasma column by means of the kink mode,
for $ t > 3 t_A$ the current sheet along $z$ breaks and forms plumes that grow and expand away from the central axis in the $xz$ and $yz$ planes.
This can be seen also in the central panels of Fig. \ref{Fig:3DFF}, that shows the 2D slices on the $yz$ plane (at $x = 0$) of the current density.
To discriminate whether these features arise from  the effect of the pressure-driven modes or the current driven modes,
we estimated the value of $\chi =$ cos$^2(\theta)$ where $\theta$ is the angle between the current density
and magnetic field vectors. For current-driven modes cos$^2(\theta) = 1$, since currents and magnetic fields
are parallel, while its value is zero for pressure-driven modes.
Figure \ref{Fig:3DFF_sliceY_ETA}, where $\chi$ for the FF-0.2 case at $t = 6t_A$ is shown,
exhibits many regions where cos$^2(\theta) = 0$, implying that the plumes may originate
from a secondary pressure-driven instability.
The development of these features results in \ess{the fragmentation} of the current sheet in the $xy$ plane in several small current sheets.
This can be seen in the lower panels of Fig. \ref{Fig:3DFF} where the 2D slices on the $xy$ plane (at $z = -1$) of the current density is shown.
Such a fragmentation ultimately results in formation of several randomly oriented filaments representing the onset of turbulent reconnection.

We show in Fig. \ref{Fig:DissipationFF} the temporal evolution of the volume averaged magnetic energy 
 (normalized to its initial value) for case FF-0.2 with $S = 2.4 \times 10^4$ (solid line),
case FF-0.2 with $S = 6.6 \times 10^3$ (dash-dotted line), case FF-10 with $S = 2.4 \times 10^4$ (dashed line), and case FF-0.2-NS with $S = 2.4 \times 10^4$ (dotted line).
The dissipation of magnetic energy during the early stages of the simulation clearly depends on the Lundquist number, as can be
seen comparing the two FF-0.2 cases with different $S$. The case without field inversion shows a negligible decrease of the magnetic energy, likely due
to ohmic dissipation.
All the configurations feature a sharp change in the decay rate
between $4 t_A$ and $10 t_A$. This interval corresponds to
the time when the CDI comes into play forming kinks in the plasma column. The \ess{decrease} of magnetic energy
in this interval can therefore be interpreted as conversion of magnetic energy into kinetic energy due to the onset of the kink instability.
For $t > 10 t_A$ the magnetic energy decreases rapidly for both the FF-0.2 configurations with current sheet, with a rate that does not seem to depend on $S$.
On the other hand, for the case without magnetic shear, the magnetic energy remains flat until $t \sim 20 t_A$.

In analogy with \S \ref{Par:Polar2D} and \S \ref{sec:PB}, we define two different phases in the temporal
evolution of the magnetic energy: phase I that starts
at the beginning of the simulation until the time where the features of the plasma column instability set in,
and phase II that begins after these instabilities sets in. The two phases are indicated by the dashed vertical lines on the left panel of Fig. \ref{Fig:DissipationFF}.
We compute the decay rates in phase I and phase II for different $S$ values for case FF-0.2, and show the
\rev{dissipation rate} obtained in such a way on the right panel of Fig. \ref{Fig:DissipationFF}.
As for the 2D and PB cases, the \rev{dissipation rate} in phase I follows the Sweet-Parker scaling (circles),
while in phase II (triangles) the slope is nearly independent on $S$ (a linear fit yields a slope $ \simeq 0.1$).
\ess{The rate of dissipation of magnetic energy during phase II is $\sim 0.1 t_A^{-1}$.}
In analogy with the 2D and 3D-PB cases, we note that this ``fast reconnection'' regime sets in after the fragmentation of the layer.

\ess{Finally, case FF-10 shows a negligible decay of magnetic energy along the whole simulation, due to the fact that the higher pitch has 
a  stabilizing effect on the 3D instabilities acting in this configuration.}

\section{Summary and Discussion}\label{Discussion}
%
%
%
We studied magnetic reconnection using three-dimensional resistive MHD simulations of a magnetically confined cylindrical plasma column featuring a circular current sheet.
Different equilibrium conditions, including radial pressure balance and a force-free field, have been considered.
Results have been compared with 2-dimensional simulation of a circular current sheet.

Our 2D simulations generalize previous studies of planar current sheets to the cylindrical case. 
The main results from these simulations are listed below :
\begin{itemize}
 \item At early stages (phase I), the magnetic dissipation rate in the current ring agrees with the Sweet-Parker scaling of $S^{-0.5}$.
 \item At later times (phase II) and for values of $S \gtrsim S_c \simeq 1 \times 10^4$, the current sheet is subjected to secondary tearing instability whereby continuous formation of plasmoids is observed. 
 The formation of plasmoids leads to the fragmentation of the initial circular sheet into multiple small-sized current sheets.
 During this stage, the decay rate increases sharply, and becomes independent of $S$, revealing the transition to a regime of fast reconnection.
 \item Eventually, the continuous formation and merging of plasmoids results in the random orientation of fragmented current-sheets that closely resemble the turbulent reconnection
  described by, e.g., \cite{Kowal2009}, \cite{Loureiro2009} and \cite{Takamoto2015}.
 \item The rate of dissipation of magnetic energy during the fast reconnection regime is \rev{$\sim 0.1 t_A^{-1}$}, consistent with previous numerical results of 2D reconnection.
\end{itemize}

In the three-dimensional case, our results can be summarized as follows:
\begin{itemize}
\item Similar to the 2D runs, the magnetic energy (during the initial phase) is dissipated
 at a rate which is consistent with Sweet-Parker scaling, $S^{-0.5}$.
\item At later times the plasma column becomes unstable to either pressure-driven or 
 current-driven instabilities (depending on the initial equilibrium configuration), the onset of which does not depend on the Lundquist number.
 In runs with same set of parameters ($\beta = 10$ and $P = 0$), the 3D pressure-driven instability starts before the 2D secondary tearing mode.   
The growth of these instabilities causes the fragmentation of the original current ring into smaller secondary current sheets (see Fig. \ref{Fig:3DPB} and Fig. \ref{Fig:3DFF}).
\item \rev{At this time an increased magnetic dissipation is observed (phase II).} The \rev{dissipation rate} becomes independent 
of $S$ and is of the order of \rev{$(\sim 0.1 \-- 0.5) t_A^{-1}$}
(see Fig. \ref{Fig:DissipationPB} and Fig. \ref{Fig:DissipationFF}). 
\item The \rev{dissipation rate} starts to become independent of $S$ for $S \simeq 10^{-3}$, a threshold value which is an order of magnitude smaller than the one obtained from 2D runs.
\end{itemize}
\rev{We point out that the dissipation rates reported here result from the interplay between magnetic reconnection and the turbulence induced by the instabilities arising in each configurations.
This may lead to energy dissipation rates that are faster than the actual reconnection rate and
could explain the differences between our findings ($\gtrsim 0.1 t_A^{-1}$) and the results reported in previous reference studies ($\sim 0.01 t_A^{-1}$).
On the other hand, three-dimensional simulations without magnetic shear, that are not expected to develop magnetic reconnection, 
do not show relevant dissipation. 
In summary, we find that the 3D instabilities alone dissipate the magnetic energy inefficiently. 
However, they play a major role in enhancing the rate of magnetic dissipation in presence of reconnection.
}

We emphasize that the Lundquist numbers for the above 3D simulations lie in the range $10^3\--10^4$ and no formation of secondary tearing instability is observed. 
The ``fast reconnection'' regime is, therefore, a mere effect of the 3D instabilities.

A similar effect was reported in recent 3D simulations by \cite{Oishi2015},
where they attributed the early fast reconnection regime to an unspecified 3D instability.
Our detailed analysis obtains consistent results in a different configuration (magnetically confined plasma column) and provides clear evidence that the onset of ``fast reconnection''
is triggered by well-known plasma instabilities (pressure- or current-driven).

Our results can be relevant in the context of MHD jets, where these instabilities are likely to operate. 
Typical astrophysical environments are active galactic nuclei, microquasars and pulsar wind nebulae.
Here, magnetic reconnection has been recently invoked as an efficient mechanism to accelerate particles to non-thermal energies \citep{Sironi2014,deGouveia2015} up to PeV energies \citep{Cerutti2013}.
Plasma instabilities in jets, therefore, could trigger fast magnetic reconnection
episodes \citep{Lyubarsky2012,Giannios2013} that may account for the observed fast variability and non-thermal features in these astrophysical scenarios,
like, e.g., the $\gamma$-ray flares from the Crab Nebula \citep{Tavani2011,Striani2011}, or the very rapid variability, $\sim$ 10 min,
detected, e.g., in PKS 2155 \citep{Aharonian2007} and PKS 1222 \citep{Aleksic2011}.
Our results can, however, be applied only in the reference frame of the jet as no velocity shear has been considered. 
Besides, a more detailed analysis would require direct investigation of particle acceleration.
These issues will be explored in forthcoming studies.

\section*{Acknowledgement}

We thank an anonymous referee for his/her comments. 
We acknowledge the CINECA award under the ISCRA initiative,
for the availability of high performance computing resources
and support.
%
%
%
\input{MS_Striani_Rev.bbl}


\bsp	
\label{lastpage}
\end{document}